\documentclass[epj]{svjour}
\usepackage[T1]{fontenc}
\usepackage[latin9]{inputenc}
\usepackage{color}
\usepackage{float}
\usepackage{units}
\usepackage{amsmath}
\usepackage{amssymb}
\usepackage{graphicx}
\usepackage{esint}
\newcommand{\lyxdot}{.}

\usepackage{graphics}
\begin{document}
\title{Spatiotemporal correlations between plastic events in the shear flow of
athermal amorphous solids}
%\subtitle{Do you have a subtitle?\\ If so, write it here}
\author{Alexandre Nicolas\inst{1} \and J\"org Rottler \inst{2} \and Jean-Louis Barrat \inst{1,}\inst{3}% etc
% \thanks is optional - remove next line if not needed
%\thanks{\emph{Present address:} Insert the address here if needed}%
}                     % Do not remove
\offprints{}          % Insert a name or remove this line
\institute{Univ. Grenoble 1/CNRS, LIPhy UMR 5588, Grenoble, F-38041, France
 \and Department of Physics and Astronomy, The University of British Columbia,6224 Agricultural Road, Vancouver, British Columbia V6T 1Z4, Canada
  \and Institut Laue-Langevin, 6 rue Jules Horowitz, BP 156, F-38042 Grenoble, France}
\date{Received: date / Revised version: date}
% The correct dates will be entered by Springer
%
\abstract{
The slow flow of amorphous solids exhibits striking heterogeneities:
swift localised particle rearrangements take place in the midst of
a more or less homogeneously deforming medium. Recently, experimental
as well as numerical work has revealed spatial correlations between
these flow heterogeneities. Here, we use molecular dynamics (MD) simulations
to characterise the rearrangements and systematically probe their correlations both
in time and in space. In particular, these correlations display a
four-fold azimuthal symmetry characteristic of shear stress redistribution
in an elastic medium and we unambiguously detect their increase in range with time. With increasing shear rate, correlations become shorter-ranged and more isotropic.
In addition, we study a coarse-grained model motivated by the observed
flow characteristics and challenge its predictions directly with the MD simulations. While the model captures both macroscopic and local properties rather satisfactorily, the agreement with respect to the spatiotemporal correlations is at most qualitative. The discrepancies provide important insight into relevant physics that is missing in all related coarse-grained models that have been developed for the flow of amorphous materials so far, namely the finite shear wave velocity and the impact of elastic heterogeneities on stress redistribution. 
 % end of PACS codes
} %end of abstract
\titlerunning {Spatiotemporal correlations of plasticity in flowing amorphous solids}
\maketitle

\section{Introduction}

When a simple liquid is sheared, it flows homogeneously, and its flow
is traditionally viewed as a uniform slide of vanishingly thin layers
of fluids past each other. On the other hand, if shear is applied
to an amorphous solid, the response of the material is highly heterogeneous,
in that small regions rearrange rapidly while the rest of the material
responds elastically, in a more or less affine way \cite{Argon1979,Schall2007}. In extreme cases,
the material may fracture \cite{Shimizu2006,Furukawa2009b,Leocmach2014,Bonamy2011};
the shear strain is then entirely borne by a thin layer of matter
which has lost its internal cohesion. Material fracture is the most
acute case of shear localisation, whereby the deformation is localised
in one region of the system. The occurrence of this phenomenon rules
out the study of the flow from a homogeneous perspective. But, even
when sheared amorphous solids do not exhibit permanent shear localisation,
there is growing evidence of the existence of correlations between
the localised rearranging regions (referred to as plastic events in
the following), that is, of a spatial organisation of the flow at
intermediate time scales: In Ref.\cite{Chikkadi2012a}, Chikkadi and co-workers observed a 
colloidal glass with confocal microscopy and demonstrated that the non-affine displacements in the material were spatially correlated, prior to
any shear-banding instability, while 
Mandal, Varnik, and colleagues \cite{Mandal2013b,Varnik2014} supported such experimental findings with numerical simulations.
The observed correlations are interpreted as the effect of the long-range
elastic deformation field induced by a plastic event in the material,
at the origin of avalanches of plastic events. In a uniform linear
elastic medium, this field is quadrupolar\cite{Eshelby1957,Picard2004}, viz., $\mathcal{G}\left(r\right)\sim\cos\left(4\theta\right)/r^2$
for a plastic event occurring at the origin, in two dimensions.

Predicting exactly where the next plastic event will occur in the
material is an intricate task that is bound to depend sensitively
on detailed knowledge of the current, static configuration of the
system\cite{Widmer-Cooper2008,Tsamados2009,Brito2009,Tanguy2010,Manning2011}. Alternatively, one may choose to investigate to what extent
the position of the next plastic event is influenced by that of its
predecessors, in the hope that extensive information about the dynamical
organization of the flow will thus be revealed. The characterisation of such correlations between plastic events is the objective of this work. Although our tools will be slightly different, we note that similar studies have appeared in two recent publications. In \cite{Chattoraj2013}, long-lived correlations of  the local strain field observed in molecular dynamics (MD) simulations were taken as evidence of the  importance of localised plastic events in  a flowing  liquid. In \cite{Benzi2014}, similar correlations were observed in a numerical model of a dense emulsion undergoing shear flow between solid plates.

In this work,  we will concentrate on a numerical study of  the flow of a very simple amorphous solid in the athermal
limit. We will propose a detailed description of
the plastic events and their dynamical correlations, resolved both in
space \emph{and} time. The influence of the applied shear rate is
studied. In order to ascertain the origin of the prominent features
of the correlations, we investigate a coarse-grained model closely
connected to the observed flow phenomenology in complement to the
atomistic simulations.

The article is structured as follows: In Section \ref{sec:Atomistic_sim_methodology},
we provide the reader with the technical details pertaining to the
MD simulations, and present the observable that
will be used to measure the local rearrangements. In Section \ref{sec:Macroscopic_rheology},
we report the general properties of the simulated flow, with a particular
focus on the statistics of individual plastic events. On the basis
of these observations, a coarse-grained model is presented in Section
\ref{sec:Coarse_grained_model} and the general agreement of the model
with the atomistic simulations is immediately assessed. Finally, Section
\ref{sec:Correlations} is dedicated to our main contribution, namely,
a detailed study of the spatiotemporal correlations between successive
plastic events and the interpretation of their salient features.

\section{Atomistic simulations at zero temperature\label{sec:Atomistic_sim_methodology}}

To probe the flow properties of amorphous solids, we resort to MD
simulations of an amorphous system under shear. More precisely, we simulate
a binary mixture of A and B particles, with $N_{A}=32500$ and $N_{B}=17500$,
of respective diameters $\sigma_{AA}=1.0$ and $\sigma_{BB}=0.88$,
confined in a square box of dimensions $205\sigma_{AA}\times205\sigma_{AA}$, with periodic
boundary conditions. The system is at reduced density 1.2. The particles, of mass $m=1$,
interact via a pairwise Lennard-Jones potential,
\begin{equation}
V_{\alpha\beta}\left(r\right)=4\epsilon_{\alpha\beta}\left[\left(\frac{\sigma_{\alpha\beta}}{r}\right)^{12}-\left(\frac{\sigma_{\alpha\beta}}{r}\right)^{6}\right],
\end{equation}
 where $\alpha,\beta=A,\, B$, $\sigma_{AB}=0.8$,$\epsilon_{AA}=1.0$,~$\epsilon_{AB}=1.5$,
and $\epsilon_{BB}=0.5$. The potential is truncated at $r=2.5\sigma_{AA}$ and shifted for continuity. Simple shear $\gamma$ is imposed at rate $\dot\gamma$
by deforming the box dimensions and remapping the particle positions. 

We conduct our study in the athermal limit, by thermostating the
system to a zero-temperature, so that no fluctuating force appears
in the equations of motion, \emph{viz.,}
\begin{eqnarray}
\frac{d\mathbf{r_{i}}}{dt} & =\mathbf{p_{i}}/m\cr
\frac{d\mathbf{p_{i}}}{dt} & =-\sum_{i\neq j}\frac{\partial V\left(\mathbf{r_{ij}}\right)}{\partial\mathbf{r_{ij}}}-\mathbf{p_{i}}/\tau_d
\label{eq:eq_of_motion_MD}
\end{eqnarray}
where $\left(\mathbf{p_{i}},\mathbf{r_{i}}\right)$ are the momentum
and position of particle $i$ in the deforming frame.

Besides the interparticle forces, the
motion of particle \emph{i} is subject to a damping force $-\mathbf{p_{i}}/\tau_d$,
that models friction against solvent molecules in a mean-field way.
Here, $\tau_d=1$ is the Langevin damping time. The relevance of this
specific implementation of friction shall be discussed in Section
\ref{sub:Successes-and-limitations}. Equations \ref{eq:eq_of_motion_MD}
are integrated with the velocity Verlet algorithm
with $\delta t=0.005$. In all the following, we use $\tau_{LJ}\equiv \sqrt{m\sigma_{AA}^2 / \epsilon}$ as the unit of time and $\sigma_{AA}$
as the unit of length.

To obtain the initial glassy states, we quenched the system at constant
volume from the liquid state down to zero temperature at a fast rate.
Note that, before any data were collected, the system was always pre-sheared
for $\gamma=0.2$ to ensure that the steady state had been reached.

\section{Macroscopic rheology \& Statistics of plastic events \label{sec:Macroscopic_rheology}}

In this section, we analyse the global rheology of the system and
collect evidence in support of the general scenario of plastic events embedded in an elastic medium outlined in the
introduction. We will
also characterize plastic events at a statistical level.

\subsection{Flow curve}

\begin{figure}[ht]
\begin{centering}
\includegraphics[width=7cm]{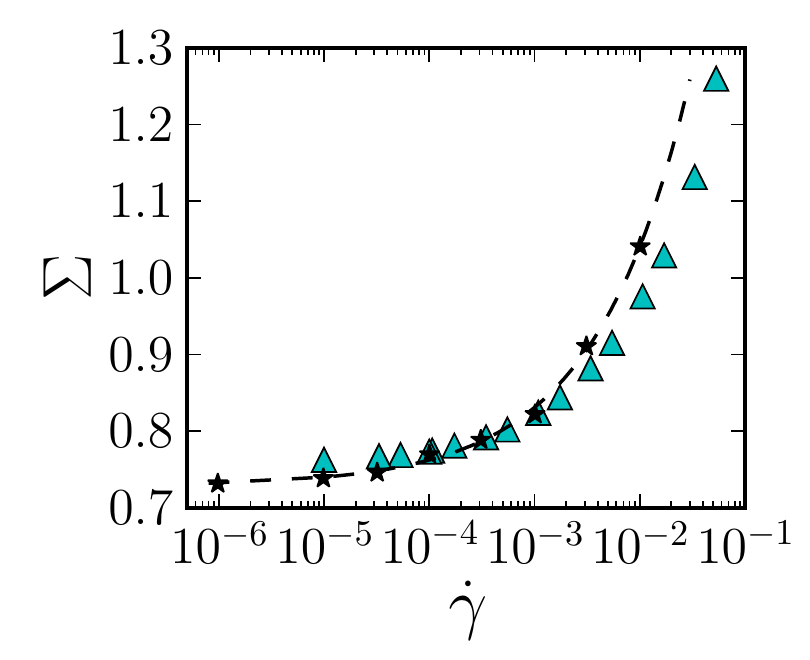}
\par
\end{centering}
\caption{\label{fig:atomistic_flow_curve}Dependence of the macroscopic shear
stress $\Sigma$ on the applied shear rate $\dot{\gamma}$. (\emph{Black
stars}) MD simulation; (\emph{blue triangles}) coarse-grained model, with $N=64\times64$ blocks.
The dashed black line is a fit to the Herschel-Bulkley equation, $\Sigma=0.73+2.9\dot{\gamma}^{0.48}$.}
\end{figure}

The dependence of the macroscopic shear stress $\Sigma$ on the applied
shear rate is shown in Fig.\ref{fig:atomistic_flow_curve}; it is
well described by the Herschel-Bulkley law $\Sigma=0.73+2.9\dot{\gamma}^{0.48}$. 
Regarding the bulk mechanical properties of the system, plotting the stress
as a function of strain at a given shear rate yields a shear modulus
$\mu\simeq17$ for the system (prior to deformation) and a macroscopic
yield strain $\gamma_{y}$ of order 5-10\%.

\subsection{Stress autocorrelation function}

\begin{figure}[ht]
\begin{centering}
%\subfloat[MD simulations]{\begin{centering}
\includegraphics[width=7cm]{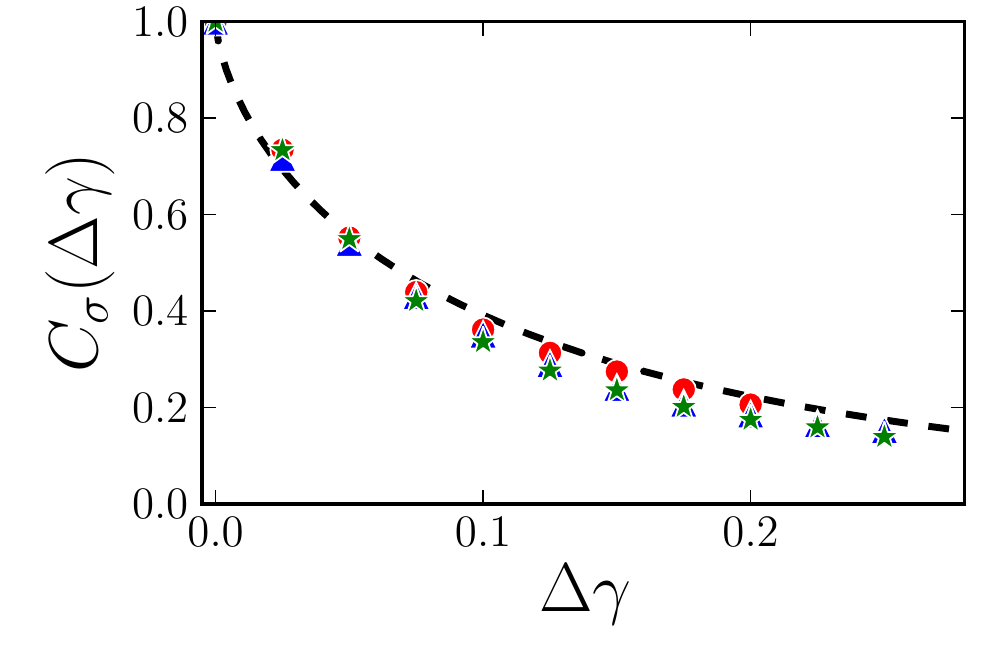}
%\par\end{centering}
\par\end{centering}
\begin{centering}
%\subfloat[Coarse-grained model]{\begin{centering}
\includegraphics[width=7cm]{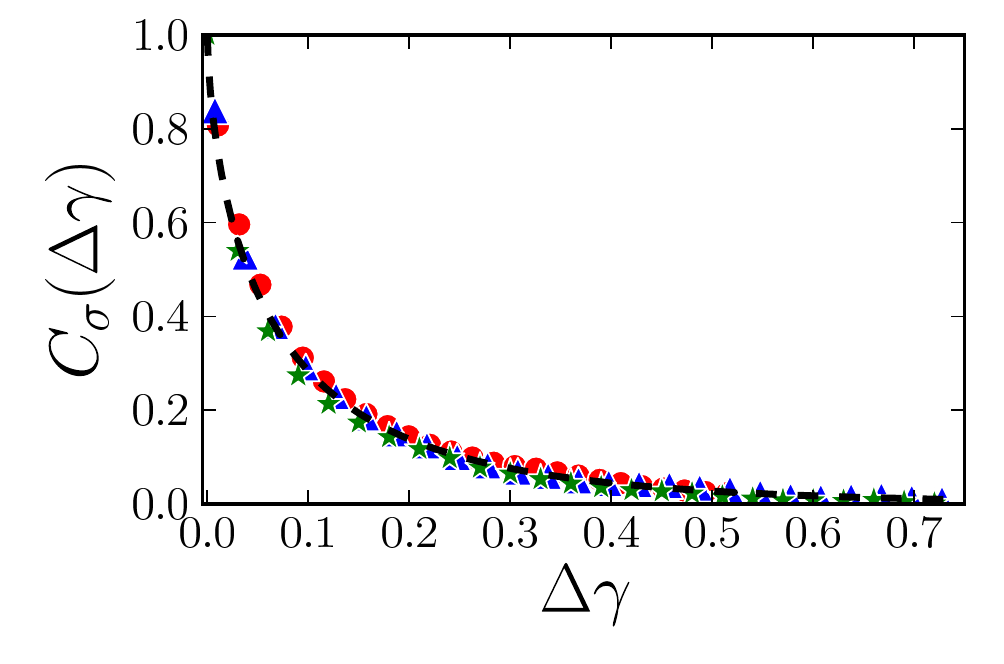}
%\par\end{centering}
\par\end{centering}

\caption{Autocorrelation function $C_{\sigma}\left(\Delta\gamma\right)$ of
the local shear stress fluctuations at applied shear rates (\emph{red
dots}) $\dot{\gamma}=10^{-5}$, (\emph{blue triangles}) $\dot{\gamma}=10^{-4}$,
and (\emph{green stars}) $\dot{\gamma}=10^{-3}$. Top panel:
results from MD simulations. Bottom panel: results from the coarse-grained model. The
dashed lines represent a fitting to a stretched exponential $C_{\sigma}\left(\Delta\gamma\right)=\exp\left[\left(\frac{-\Delta\gamma}{\Delta\gamma^{\star}}\right)^{\beta}\right]$,
with $\left(\beta=0.68,\,\Delta\gamma^{\star}=0.11\right)$ for the
MD data and $\left(\beta=0.65,\,\Delta\gamma^{\star}=0.07\right)$
for the coarse-grained results. \label{fig:stress_autocorrelation}}
\end{figure}

Turning to more local quantities, in Fig.\ref{fig:stress_autocorrelation}(top panel)
we plot the autocorrelation function 

\begin{equation}
C_{\sigma}\left(\Delta\gamma\right)\equiv\frac{\left\langle \delta\sigma_{xy}\left(\gamma\right)\delta\sigma_{xy}\left(\gamma+\Delta\gamma\right)\right\rangle }{\left\langle \delta\sigma_{xy}^{2}\right\rangle }
\end{equation}
of the local shear stress fluctuations $\delta\sigma_{xy}\equiv\sigma_{xy}-\left\langle \sigma_{xy}\right\rangle $
experienced by each particle. The averages are performed over time. We observe a nice collapse of
the data for the different shear rates. This confirms that the applied
strain $\Delta\gamma$, and not the absolute time $t$, causes the
decorrelation in this driven athermal system, in line with the idea
of periods of elastic accumulation of stress interspersed with shear-induced
plastic events. The master curve is fairly well fit by a stretched
exponential $\exp\left[-\left(\frac{\Delta\gamma}{\Delta\gamma^{\star}}\right)^{\beta}\right]$,
with an exponent $\beta=0.68$ and a characteristic strain $\Delta\gamma^{\star}=0.11$
close to the macroscopic yield strain.

\subsection{Indicator of plastic activity\label{sec:PlasticActivity}}

\begin{figure}[ht]
\begin{centering}
\includegraphics[width=7cm]{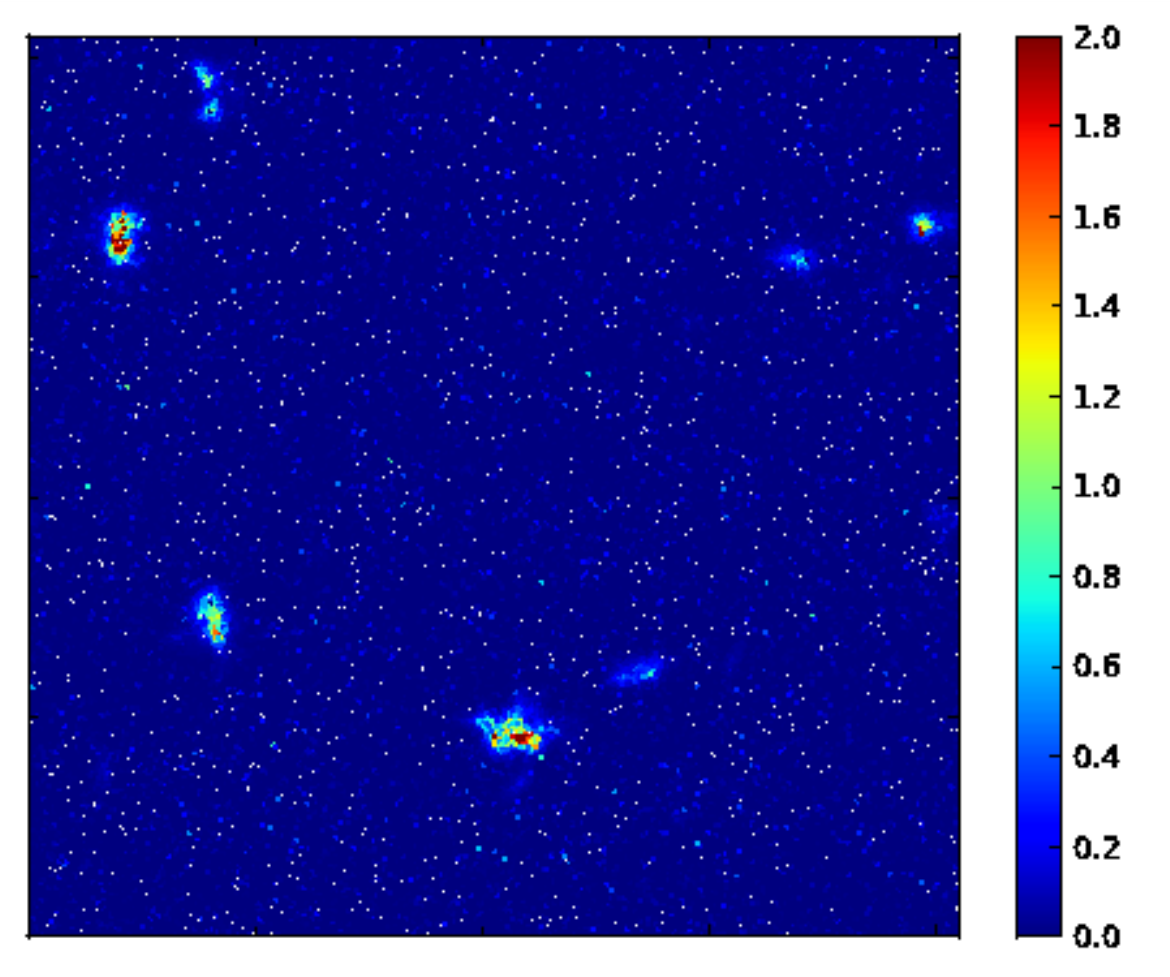}
\par\end{centering}

\caption{Snapshot of the $D_{min}^{2}$ field at an applied shear rate $\dot{\gamma}=10^{-4}$.
\label{fig:D2min_snapshot-1}}
\end{figure}

Let us now focus on plastic events. In order to detect them, we make
use of the $D_{min}^{2}$ quantity presented by Falk and Langer in Ref.
\cite{Falk1998}, which evaluates deviations from an affine deformation
on a local scale. This quantity has been used with noted success to characterize plasticity
\cite{Falk1998,Chikkadi2011,Chikkadi2012a,Chikkadi2012b,Mandal2013b,Varnik2014,Keim2014}; in particular, it was shown to 
yield results consistent with other measures of nonaffinity in Ref. \cite{Chikkadi2012b}.
$D_{min}^{2}$ is defined locally, around a particle labelled \emph{i},
as the minimum over all possible linear deformation tensors $\boldsymbol{\epsilon}$
of
\begin{eqnarray*}
D^{2}\left(i;t,\delta t\right) & = & \sum_{j}\left[r_{ij}\left(t+\delta t\right)-\left(\boldsymbol{\mathbb{I}}+\boldsymbol{\epsilon}\right)\cdot r_{ij}\left(t\right)\right]^{2},
\end{eqnarray*}
where the sum runs over all neighbours $j$ of $i$, and $\boldsymbol{\mathbb{I}}$
denotes the identity matrix. The value of the time lag $\delta t$
was fine-tuned to provide a good signal over noise ratio while still
being short enough to allow a temporal resolution of the plastic events.
Figure \ref{fig:D2min_snapshot-1} presents a snapshot of $D_{min}^{2}$
values in the system: one clearly sees localised plastic regions embedded
in an affinely-deforming medium. To provide a more dynamical view
of the flow, short movies are available as Supplementary Material \cite{Supp-mat}, along with their counterparts for the coarse-grained model
presented in the next section.

Interestingly, the regions with large $D_{min}^{2}$ systematically
coincide with the regions exhibiting large  velocities
relative to the average solvent flow. This
 coincidence between the non-locally-affine displacement field
and the singular velocity confirms that large local energy dissipation
is the hallmark of a plastic event.

\subsection{Distribution of durations, magnitudes, and sizes of individual plastic
events}

\begin{figure}[t]
\begin{centering}
\includegraphics[width=7cm]{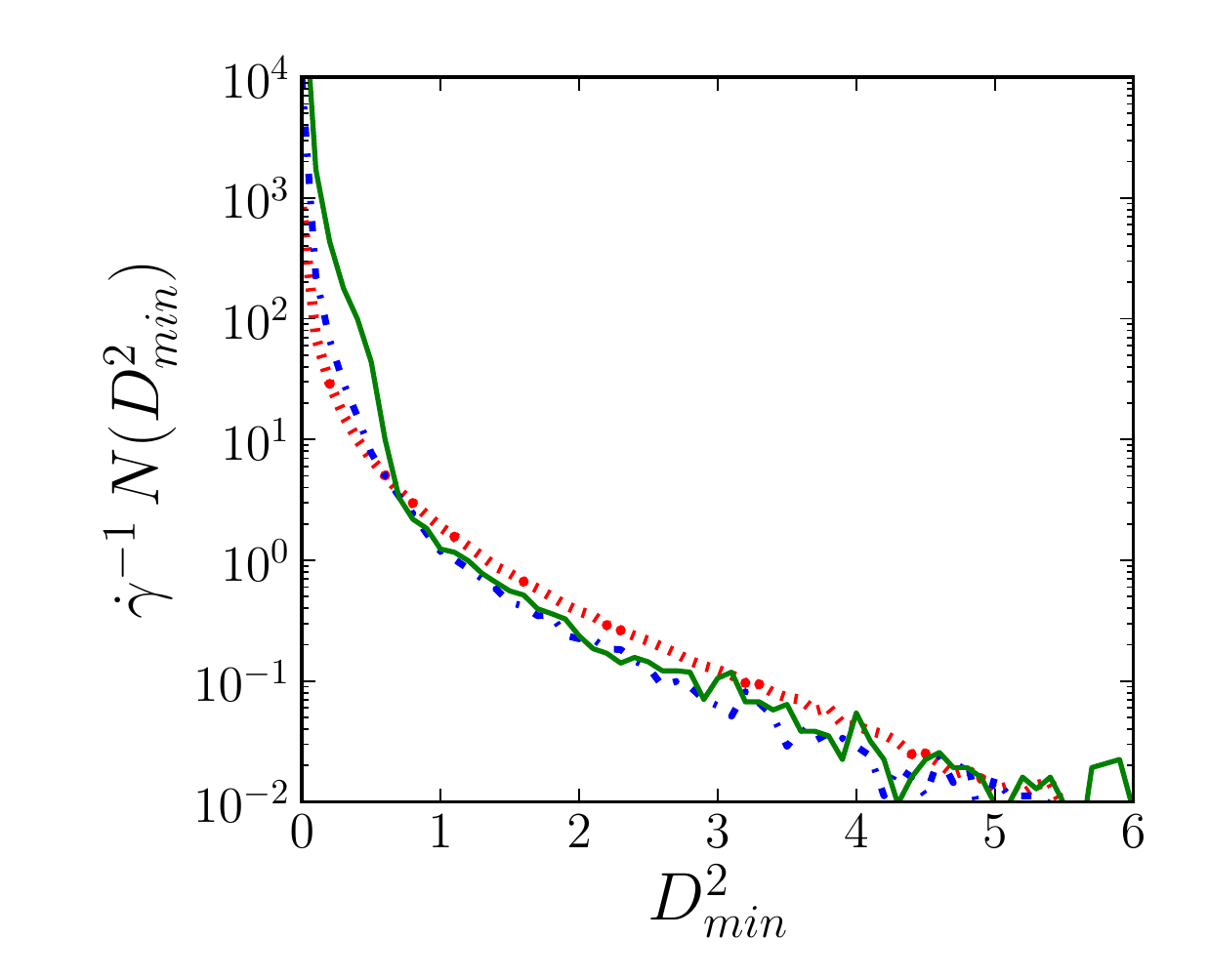}
\par\end{centering}

\caption{\label{fig:D2min_distribution}Histograms of $D_{min}^{2}$ values
at shear rates (\emph{dotted red line})$\dot{\gamma}=10^{-5}$, (\emph{dashdotted
blue line})$\dot{\gamma}=10^{-4}$, and (\emph{solid green line})$\dot{\gamma}=10^{-3}$.
The histograms collapse upon rescaling with the inverse shear rate.}
\end{figure}

\begin{figure}[t]
\begin{centering}
\includegraphics[width=7cm]{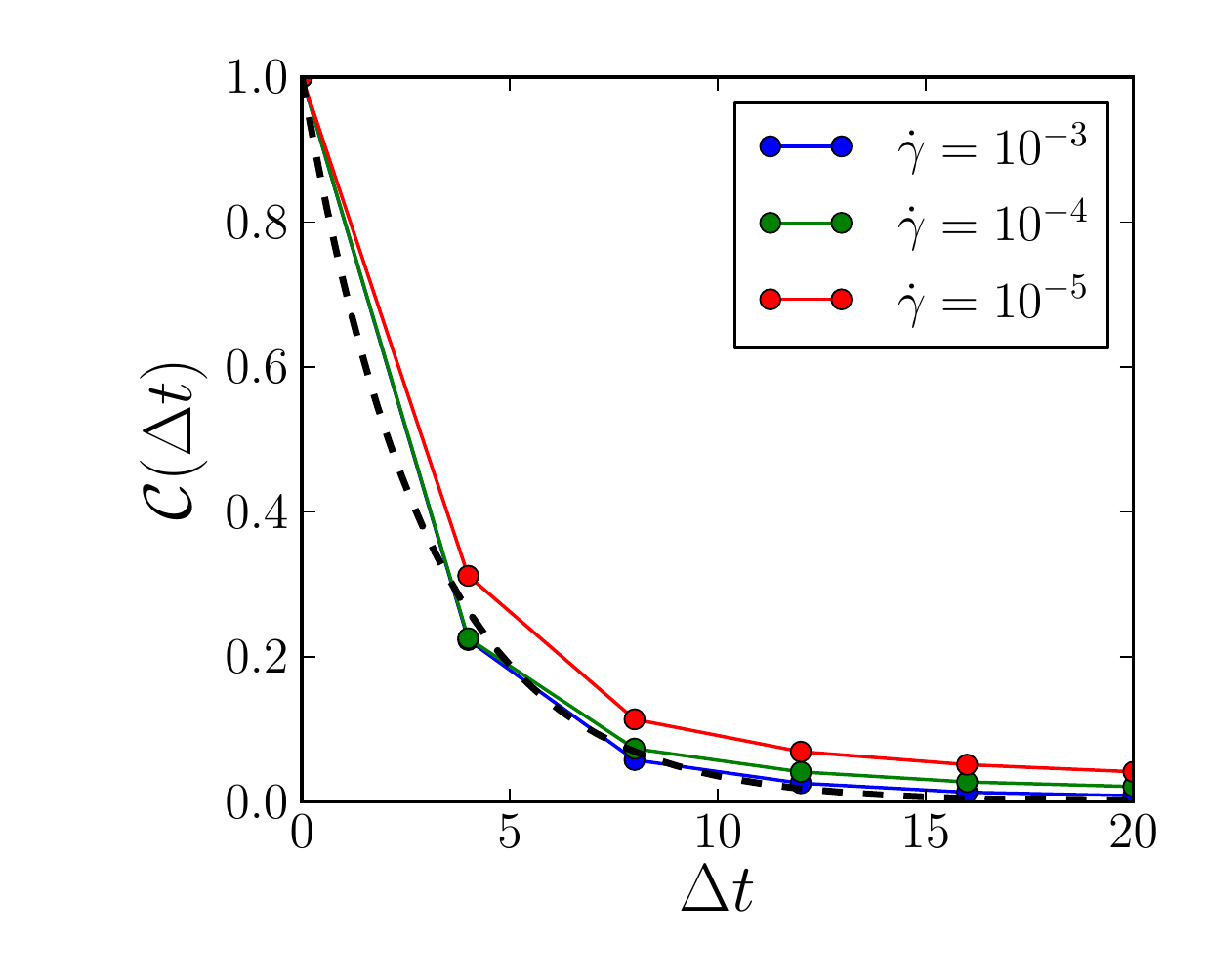}
\par\end{centering}

\caption{
\label{fig:D2min_autocorrelator}
$D_{min}^{2}$ autocorrelation function
$\mathcal{C}(\Delta t) 
\equiv
\left\langle D_{min}^{2}(r,t)D_{min}^{2}(r+\epsilon,t+\Delta t)\right\rangle 
/
\left\langle \left(D_{min}^{2}(r,t)\right)^{2}\right\rangle $
as a function of the time lag $\Delta t$, where $\epsilon \simeq 0.005 L$
has only been added to avoid numerical artifacts. The dashed black
line represents $\exp\left(\nicefrac{-\Delta t}{3.0}\right)$. 
}
\end{figure}

We now study the properties of individual plastic events in more detail.
First, by scrutinising a number of $D_{min}^{2}$ snapshots such as
the image presented in Fig.\ref{fig:D2min_snapshot-1}, we observe that
the size of plastic regions is typically a few particle diameters; this size does not depend dramatically on the shear rate.
This point will be confirmed in Section \ref{sec:Correlations} by
a detailed analysis of the spatial correlations of the $D_{min}^{2}$
field.

Further insight is gained by computing the overall distribution of
the measured $D_{min}^{2}$ values in Fig.\ref{fig:D2min_distribution}.
All distributions exhibit an exponential tail, and they furthermore collapse upon rescaling with the inverse shear rate.

Finally, the typical lifetime of a plastic event can be extracted
from the temporal decay of the $D_{min}^{2}$ autocorrelation function plotted in Fig.\ref{fig:D2min_autocorrelator}.
For the value of the damping time $\tau_d$ used in this study, it is of the order of 3 time units regardless of the shear rate.

\section{Generic coarse-grained model\label{sec:Coarse_grained_model}}

\subsection{Description of the model}

The numerical observations reported above all support the flow scenario
based on short-lived, localised, and strongly dissipative plastic
events embedded in an elastic matrix. Accordingly, we shall now introduce
a simple, 2D coarse-grained model for the rheology of athermal amorphous
solids that is motivated by these observations. 

To start with, we discretise space into a lattice $\left\{ \left(i,j\right)\right\} $
of $N=128\times128$ square-shaped elastoplastic blocks, each of the size of a rearranging
region. By default, blocks are elastic, in which case the (tensorial)
deviatoric stress $\boldsymbol{\sigma}\left(i,j\right)$ and strain
$\boldsymbol{\epsilon}\left(i,j\right)$ tensors carried by each block
obey Hooke's law, \emph{viz}.,
\begin{equation}
\left(\begin{array}{c}
\sigma_{xx}\left(i,j\right)\\
\sigma_{xy}\left(i,j\right)
\end{array}\right)=2\mu\left(\begin{array}{c}
\epsilon_{xx}\left(i,j\right)\\
\epsilon_{xy}\left(i,j\right)
\end{array}\right),\label{eq:Hooke}
\end{equation}
where $\mu$ is the shear modulus. Here, we have postulated incompressibility,
i.e., $\epsilon_{yy}\left(i,j\right)=-\epsilon_{xx}\left(i,j\right)$.
Plasticity is incorporated into the model by allowing blocks to yield
(i.e., switch to the plastic state) as soon as the following yield
criterion is fulfilled,
\begin{equation}
\left\Vert \boldsymbol{\sigma}\left(i,j\right)\right\Vert \equiv\sqrt{\sigma_{xx}^{2}\left(i,j\right)+\sigma_{xy}^{2}\left(i,j\right)}\geqslant\sigma_{y}\left(i,j\right),\label{eq:meso_yield_criterion}
\end{equation}
 where $\sigma_{y}\left(i,j\right)$ is a fixed local yield stress,
associated to an energy barrier $E_{y}\left(i,j\right)=\nicefrac{\sigma_{y}^{2}\left(i,j\right)}{4\mu}$.
Equation \ref{eq:meso_yield_criterion} is simply the well-known von Mises
yield criterion%
\footnote{Note that the von Mises and the Tresca yield criteria are equivalent
in two dimensions.%
}. Every time a block yields, the value of $E_{y}\left(i,j\right)$
is renewed; it is randomly selected from a truncated exponential distribution,
\begin{equation}
P\left(E_{y}\right)=\Theta\left(E_{y}-E_{y}^{min}\right)\exp\left(\lambda\left(E_{y}^{min}-E_{y}\right)\right),\label{eq:dist_of_yield_energies}
\end{equation}

where $\Theta$ is the Heaviside function. The coefficient $\lambda$
is chosen such that the average of the yield strain $\gamma_{y}=2\sqrt{\nicefrac{E_{y}}{\mu}}$
over $P$ coincides with the MD macroscopic yield strain, $\left\langle \gamma_{y}\right\rangle \approx 0.1$.
The introduction of a lower threshold $E_{y}^{min}$ in Eq.\ref{eq:dist_of_yield_energies}
comes down to discarding too shallow metabasins in the potential energy
landscape (PEL) of the subsystem modelled as an elastoplastic block.

A plastic block is a fluid-like inclusion embedded in an elastic region.
The implications of this fact are twofold. First, the plastic rearrangement
occurs, not instantaneously, but over a finite time scale $\tau$,
because viscous forces oppose it\cite{Nicolas2013b}. Second, the
associated distortion of the plastic region induces an additional
stress in the surrounding elastic medium\cite{Eshelby1957}. The combination
of these two effects occurring in plastic blocks, along with Eq.\ref{eq:Hooke}
for the elastic regions, leads to,
\begin{equation}
\partial_{t}\boldsymbol{\sigma}\left(i,j\right)=\mu\boldsymbol{\dot{\gamma}}+2\mu\sum_{i^{\prime},j^{\prime}}\boldsymbol{\mathcal{G}}\left(i-i^{\prime},j-j^{\prime}\right)\boldsymbol{\dot{\epsilon}^{pl}}\left(i^{\prime},j^{\prime}\right).\label{eq:evolution_eq_of_stress}
\end{equation}
Here, $\boldsymbol{\dot{\epsilon}^{pl}}\left(i^{\prime},j^{\prime}\right)=\boldsymbol{\sigma}\left(i^{\prime},j^{\prime}\right)/2\mu\tau$
if the block $\left(i^{\prime},j^{\prime}\right)$ is plastic, ${\bf 0}$
otherwise. $\tau$ is the local viscous time, and  the propagator $\boldsymbol{\mathcal{G}}$ is such
that $2\mu\boldsymbol{\mathcal{G}}\left(i-i^{\prime},j-j^{\prime}\right)\boldsymbol{\epsilon^{pl}}$
is the stress increment received by block $\left(i,j\right)$ if the
block $\left(i^{\prime},j^{\prime}\right)$ endures a plastic strain
$\boldsymbol{\epsilon^{pl}}$. Note in particular that $\boldsymbol{\mathcal{G}}\left(0,0\right)$
has negative eigenvalues, because plastic blocks relax the stress
they bear. More generally, the propagator $\boldsymbol{\mathcal{G}}$
was calculated rigorously in the limit of a pointwise inclusion in
a perfectly homogeneous elastic medium. Its expression and the way
it must be altered when simulation cell is deformed are specified
elsewhere\cite{Nicolas2014u}. Obviously, the first term
on the right hand side of Eq.\ref{eq:evolution_eq_of_stress} contains
the elastic response to the macroscopic driving and follows directly
from Eq.\ref{eq:Hooke}, while the second term deals with the effects
of plasticity. 

The swift rearrangement of particles that characterises a plastic
event corresponds to a jump between metabasins in the PEL\cite{Doliwa2003}.
In our model, it shall come to an end when a given strain $\gamma_{c}$
has been cumulated locally in the plastic phase, i.e., when 
\begin{equation}
\int\left\Vert \boldsymbol{\dot{\epsilon}}\left(t^{\prime}\right)\right\Vert dt^{\prime}\geqslant\gamma_{c},\label{eq:endPl_criterion}
\end{equation}
where the \emph{total} local deformation reads, $\ensuremath{\boldsymbol{\dot{\epsilon}}=\frac{\partial_{t}\boldsymbol{\sigma}}{2\mu}+\boldsymbol{\dot{\epsilon}^{pl}}}$.
$\gamma_{c}$ represents the typical distance between metabasins,
which clearly depends on how much the PEL has been coarse-grained,
that is, on $E_{y}^{min}$. For simplicity, we arbitrarily take $\gamma_{c}=2\sqrt{\nicefrac{E_{y}^{min}}{\mu}}$.
Following this simplification, $E_{y}^{min}$ is the \emph{only} free
parameter in the model, if one excepts the time and stress units $\tau$
and $\mu$. 

To sum up, the transitions between the elastic and plastic regimes
obey,
\begin{equation}
\text{elastic}\underset{\int_{pl}dt\left\Vert \boldsymbol{\dot{\epsilon}}\right\Vert \geqslant\gamma_{c}}{\overset{\left\Vert \boldsymbol{\sigma}\right\Vert \geqslant\sigma_{y}}{\rightleftharpoons}}\text{plastic}
\end{equation}

Finally, a coarse-grained version of convection is introduced in the
system by incrementally shifting lines of blocks in the flow direction.
For this purpose, we keep track of the exact average displacement
of each of these 'streamlines' along the flow direction, and shift
the whole line once the displacement gets larger than the size of
one block. The implementation of convection also requires to compute
the elastic propagator in a deformed frame\cite{Nicolas2014u}.

\subsection{Comparison of general features with the atomistic simulations\label{sub:Model_vs_MD_macro}}

%\subsubsection{Flow curve}

Figure \ref{fig:atomistic_flow_curve} presents a comparison of the
flow curves obtained with the coarse-grained model and with the atomistic
simulations. Note that, to allow direct comparison, the time and stress
units in the model must be specified. A reasonably good agreement between the flow curves 
is obtained by setting the shear modulus
to 12.5, a value comparable to the shear modulus of the atomistic
system prior to deformation ($\mu_{MD}=17$), and $\tau=1.5$,
which will lead to similar plastic event life times in the MD and
coarse-grained simulations. The best fits of the flow curves with Herschel-Bulkley 
equations have very similar exponents $n\simeq 0.5$.

%\subsubsection{Density of plastic regions}

To quantify the global plasticity of the system, we compute the instantaneous
surface density of plastic events, i.e., the fraction of blocks which
are plastic at a given time. In the absence of thermally-activated
plastic events, this quantity increases linearly with the shear rate,
from 0.05\% at $\dot{\gamma}=10^{-5}$ to 0.36\% at $\dot{\gamma}=10^{-4}$
and 2.8\% at $\dot{\gamma}=10^{-3}$. These values are similar to
those obtained from the atomistic simulations by integrating the tails of the $D_{min}^{2}$
distributions, in Fig.\ref{fig:D2min_distribution}, down to a reasonable (but arbitrary) lower threshold: 0.07\%, 0.3\%,
and 0.8\%, respectively.

%\subsubsection{Stress autocorrelation function}

Turning to a more local viewpoint, the autocorrelation function of
the stress fluctuations on a given block are shown in panel (b) of
Fig.\ref{fig:stress_autocorrelation}. As in the MD simulations, the
autocorrelations at different strain rates collapse onto a master
curve. Interestingly, this master curve is fitted by a stretched exponential, $\exp\left[-\left(\frac{\Delta\gamma}{\Delta\gamma^{\star}}\right)^{\beta}\right]$,
with a stretching exponent $\beta=0.65$ very close to the one used to fit
the MD data ($\beta=0.68$), although the precise value of the characteristic
strain obtained here, $\Delta\gamma^{\star}=0.07$, differs by 50\%.

%\subsubsection{Life time of a plastic event}

\begin{figure}[ht]
\begin{centering}
\includegraphics[width=7cm]{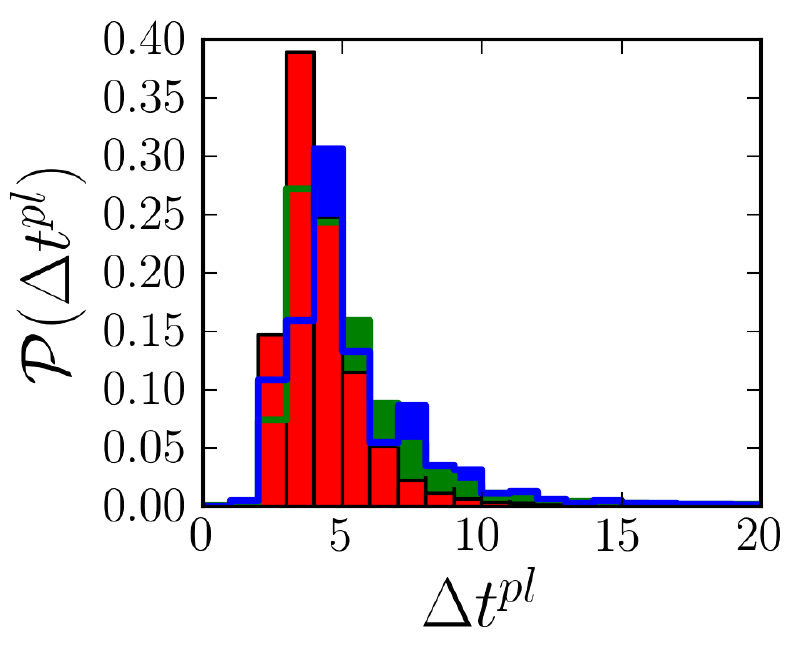}
\par\end{centering}

\caption{Histogram of the durations $\Delta t^{pl}$ of plastic events, at
(\emph{blue})$\dot{\gamma}=10^{-5}$, (\emph{green})$\dot{\gamma}=10^{-4}$,
and (\emph{red})$\dot{\gamma}=10^{-3}$ \label{fig:meso_lifetimes}.}
\end{figure}

The average life time of a single plastic event in the model is of
order a few $\tau$ (remember that we set $\tau$ to 1.5)
at all shear rates. More precisely, a noticeable decrease of the average
life time is observed as the shear rate is increased, from $8.4$ at
$\dot{\gamma}=10^{-5}$ to $4.2$ at $\dot{\gamma}=10^{-3}$. This
is not unexpected, because the criterion determining the duration
of a plastic event, Eq.\ref{eq:endPl_criterion}, involves the total
local deformation rate. Indeed, the distributions of plastic event
life times, shown in Fig.\ref{fig:meso_lifetimes}, undergo a small,
but noticeable shift to shorter times at higher shear rates.

%\subsubsection{Distribution of plastic event magnitudes}

Since we introduced a cut-off in the yield stress distribution (see Eq.\ref{eq:dist_of_yield_energies}),
the distribution of plastic event magnitudes will naturally differ
from that observed in the atomistic simulations. Nevertheless, this distribution is still roughly independent
of the applied shear rate (\emph{data not shown}).

\section{Correlations between plastic events\label{sec:Correlations}}

Having verified the agreement of the coarse-grained model with the
atomistic simulations with regard to the general flow properties, we can
now move on to the study of the correlations in the flow.

\subsection{Plastic correlator}

The individual localised rearrangements identified in Section \ref{sec:PlasticActivity} are not
random isolated events: in the athermal, quasi-static limit, Maloney and Lema\^{i}tre \cite{Maloney2004} showed numerically 
that they are essentially organised in strongly correlated avalanches. By investigating the transverse particle diffusivity, 
Lema\^{i}tre and colleagues then showed that these correlations 
persist at finite shear rates \cite{Lemaitre2009} and at finite temperatures \cite{Chattoraj2010,Chattoraj2011}. The spatial 
structure of these correlations was revealed by Chikkadi, Mandal, Varnik, \emph{et al.} \cite{Chikkadi2012a,Mandal2013b}; these researchers provided convincing experimental and numerical evidence that the correlations between flow heterogeneities, quantified by $D^2_{min}$, are long-ranged and all the more anisotropic as shear prevails over thermal effects, i.e., at larger Peclet numbers.
To do so, they monitored particle displacements in a driven  {}``hard sphere'' colloidal glass with confocal microscopy and were able to reproduce their
experimental observations qualitatively with MD simulations. Quantitatively, some discrepancies were found between simulations and experiments, the latter displaying longer correlations, with a power law decay in space. 

Here, we purport to extend these studies and unveil the full dynamical picture by resolving
correlations between the plastic events both in time and
in space, for different shear rates, in the athermal regime. The emphasis shall then be put on the causal links that
exist between successive plastic events. To this end, we use
the following two-time, two-point \emph{plastic correlator},
\begin{eqnarray}
\mathcal{C}_{2}\left(\Delta r,\Delta t\right) & \equiv & \alpha\Big(\left\langle \overline{D_{min}^{2}\left(r,t\right)D_{min}^{2}\left(r+\Delta r,t+\Delta t\right)}\right\rangle \\ \nonumber
 &  & -\left\langle \overline{D_{min}^{2}\left(r,t\right)}\cdot\overline{D_{min}^{2}\left(r,t+\Delta t\right)}\right\rangle \Big),
\end{eqnarray}
where the brackets denote an average over time $t$, the bars represent
an average over spatial coordinate $r$, and the prefactor 
$\alpha\equiv \Big[ \left\langle \overline{ (D_{min}^{2}(r,t))^2 }\right\rangle - \left\langle \overline{ (D_{min}^{2}(r,t)) }^2\right\rangle \Big]^{-1}$
is chosen such that $\mathcal{C}_{2}\left(\Delta r=0,\Delta t=0\right)=1$.
Clearly, $\mathcal{C}_{2}$\textbf{\emph{ }}measures the (enhanced or reduced) likelihood
that a plastic event occurs at $r+\Delta r$ if a plastic event was
active at position $r$ some prescribed time $\Delta t$ ago.\textbf{\emph{
}}In the coarse-grained simulations, a sensible equivalent is,

\begin{eqnarray}
\mathcal{C}_{2}\left(\Delta r,\Delta t\right) & \equiv & \alpha^\prime\Big(\left\langle \overline{n\left(r,t\right)n\left(r+\Delta r,t+\Delta t\right)}\right\rangle \nonumber \\
 &  & -\left\langle \overline{n\left(r,t\right)}\cdot\overline{n\left(r,t+\Delta t\right)}\right\rangle \Big),
\end{eqnarray}
where $n\left(r,t\right)=1$ if the block at position $r$ is plastic
at time $t$, $n\left(r,t\right)=0$ otherwise, and 
$\alpha^\prime \equiv \Big[ \left\langle \overline{ n(r,t)^2 }\right\rangle - \left\langle \overline{ n(r,t) }^2\right\rangle \Big]^{-1}$ is, again, a normalisation prefactor.

\subsection{Decay of the intensity of the correlation with time}

Plastic correlations naturally fade away with time, but one may wonder
whether their decay is more appropriately described in terms of the
absolute time $t$ or the strain $\gamma$. Quite interestingly, in
the atomistic simulations as well as in the coarse-grained model,
absolute time turns out to be the adequate unit of measurement, as
evidenced by comparing the evolution of the correlations at different
shear rates.

It should be pointed out that this does not conflict with the decay
of stress correlations as a function of the \emph{strain}.  Stress correlations exist during the loading phase preceding the shear transformation, whose duration is typically determined by the yield strain. On the other hand, the duration of the plastic activity phase is mostly determined by the local damping time, and is only weakly dependent on strain rate. Correlations in plastic activity are therefore expected on the time scale on a single event, or on somewhat  longer time scales in the event of correlated avalanches, but they will remain limited to finite times even for vanishingly small applied strain rates.

% The difference
%comes from the fact that plastic correlations are conditional probabilities,
%hinging on the occurrence of a first plastic event at $(r,t)$. On
%the contrary, stress correlations involve all pairs of instants separated
%by a given $\Delta t$; therefore, one generally has to wait for the
%whole elastic regime to observe a decorrelation.

\subsection{Maps of plastic correlations at various shear rates}

\begin{figure*}
\begin{centering}
\includegraphics[width=5cm]{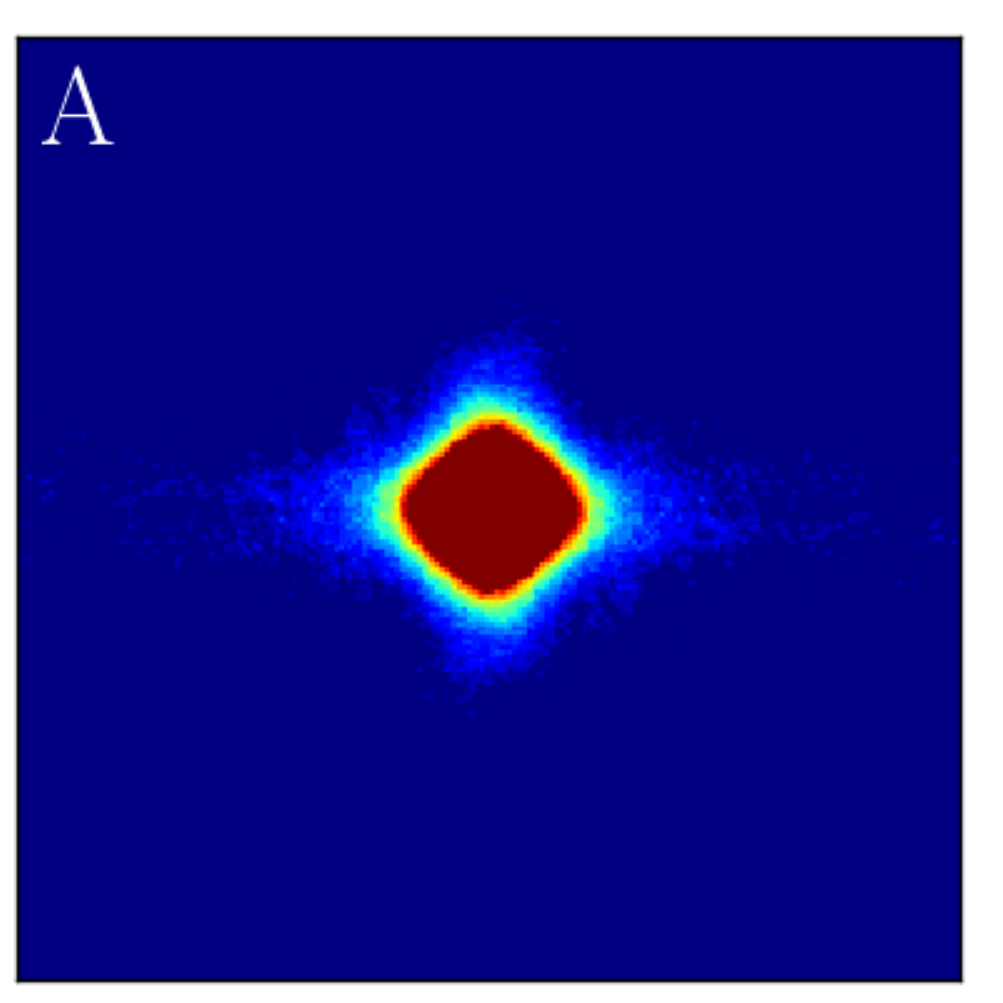}\includegraphics[width=5cm]{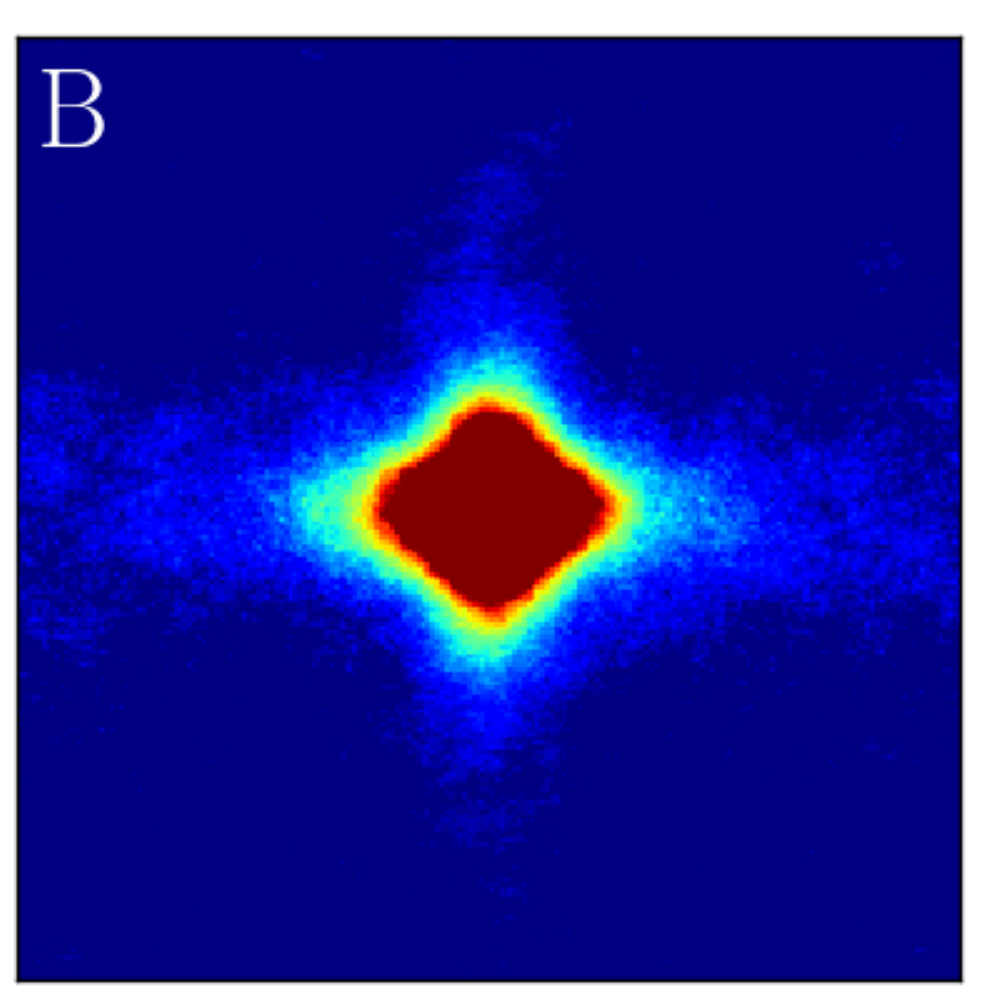}\includegraphics[width=5cm]{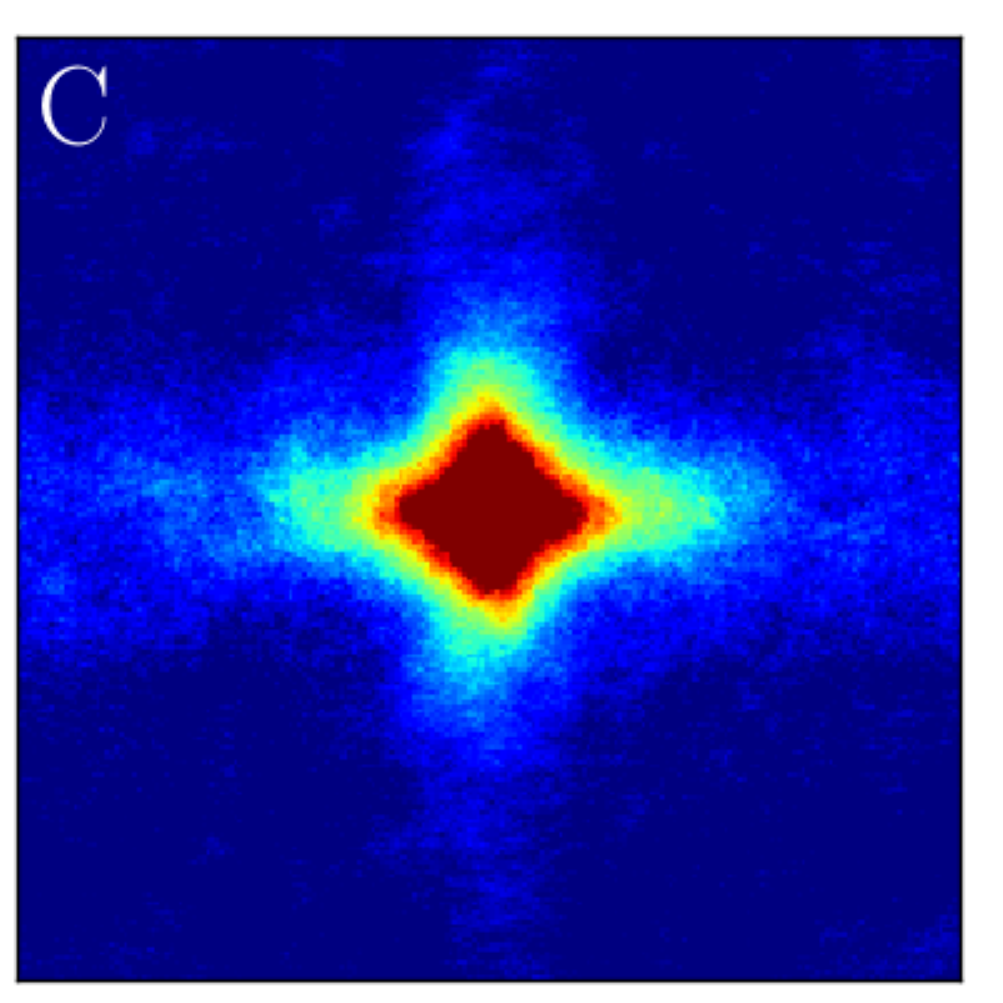}
\par\end{centering}

\begin{centering}
\includegraphics[width=5cm]{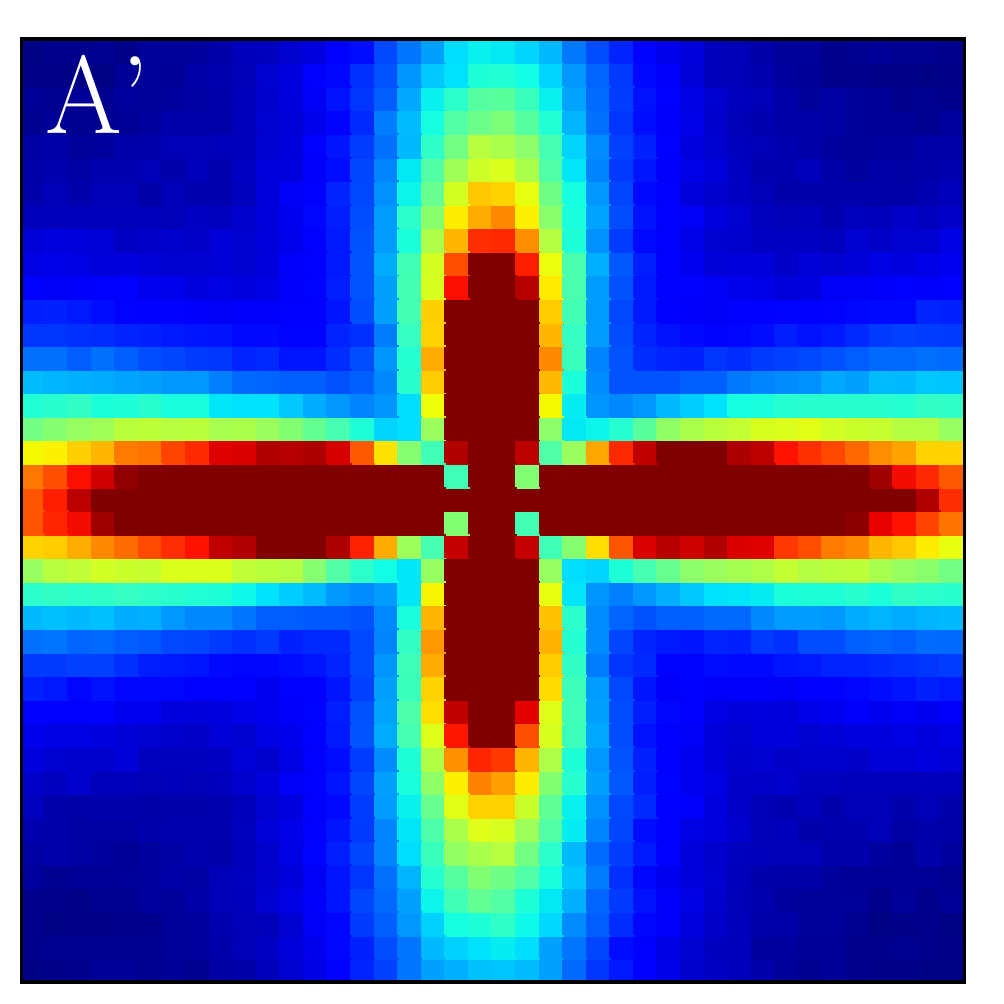}\includegraphics[width=5cm]{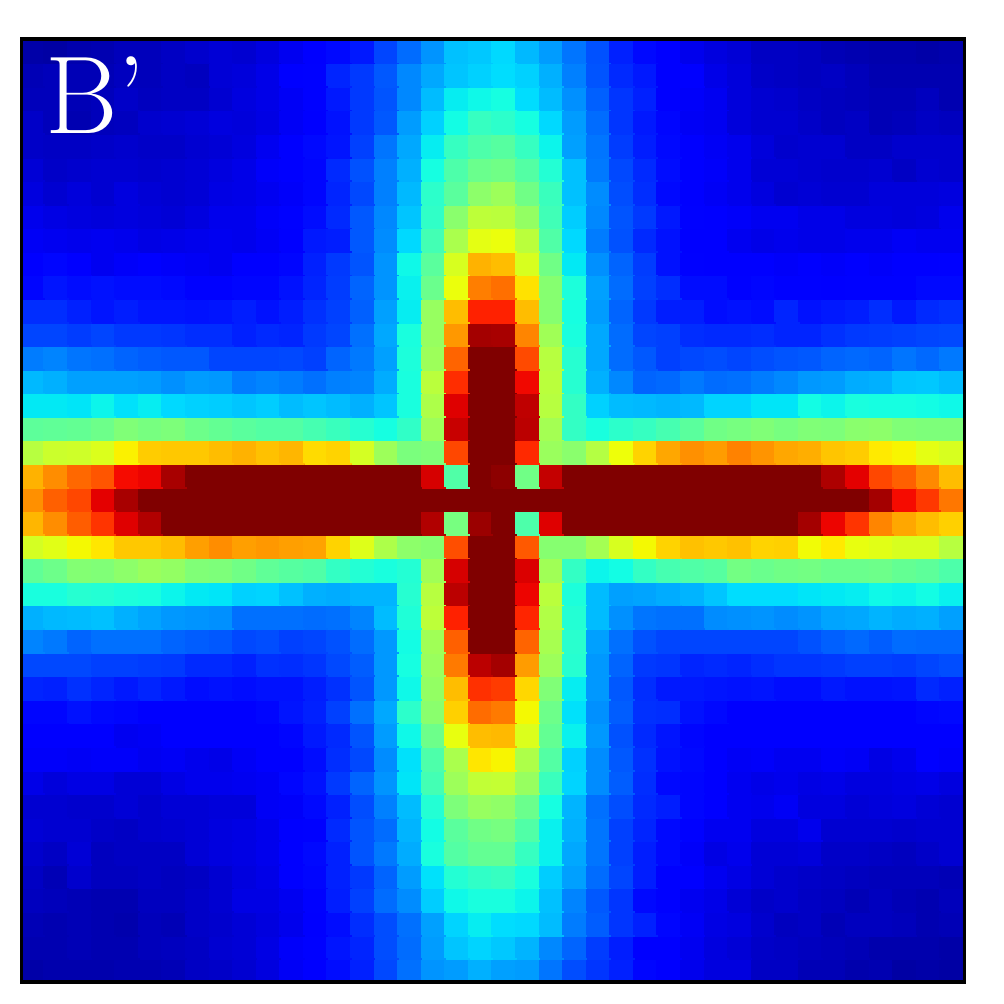}\includegraphics[width=5cm]{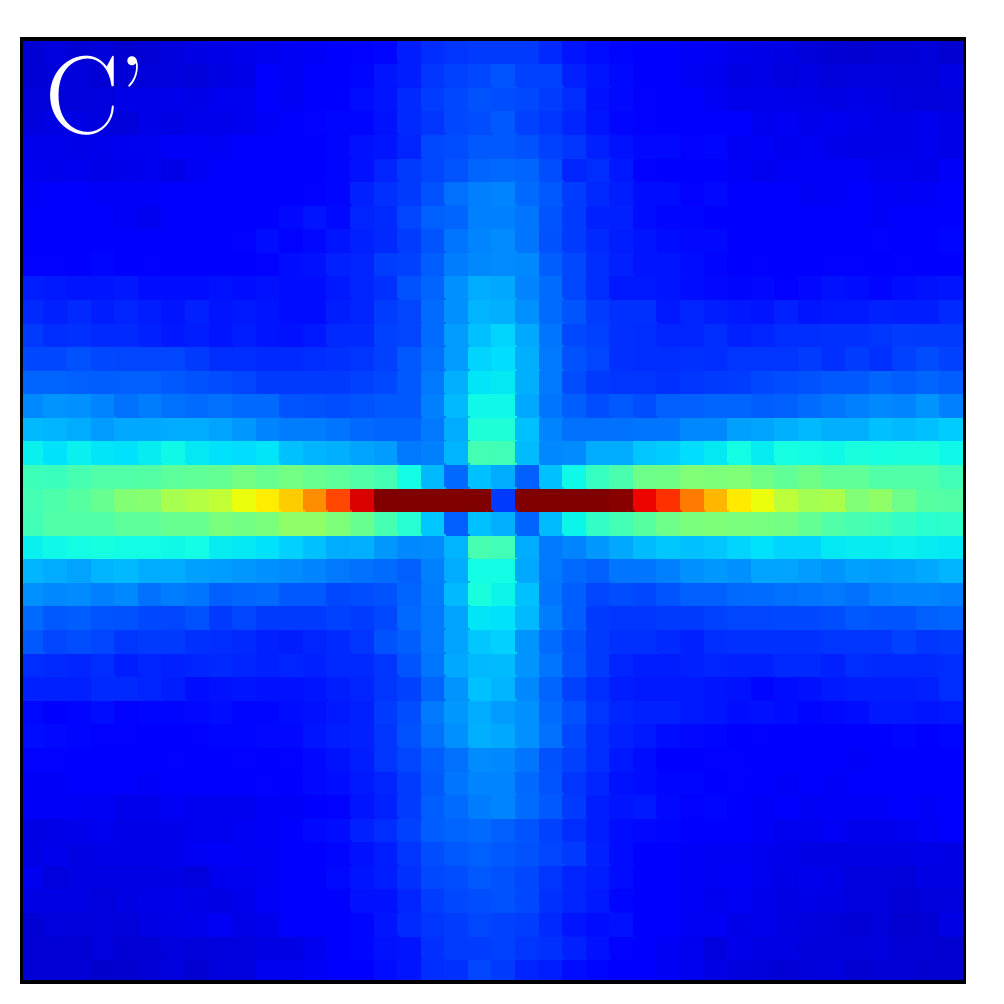}
\par\end{centering}

\caption{\label{fig:PlCorrLOW}Colour maps of the plastic correlator $\mathcal{C}_{2}$
at very low shear rate $\dot{\gamma}=10^{-5}$ for time lags (\emph{A})
$\Delta t=0$, (\emph{A'}) $\Delta t=$1, (\emph{B and B'}) $\Delta t=8$,
(\emph{C and C'}) $\Delta t=20$ . \emph{Top row: }MD simulations.
\emph{Bottom row:} coarse-grained model. Note that for the coarse-grained
model, we restrict the view to a region of the size (40x40 blocks)
of the MD simulation cell. The colour code ranges from dark blue,
for values below $-5\cdot10^{-4}$, to dark red, for all values $\geqslant5\cdot10^{-3}$. Note that the largest values are considerably greater
than the chosen upper cutoff, $5\cdot10^{-3}$. }
\end{figure*}

\begin{figure*}
\begin{centering}
\includegraphics[width=5cm]{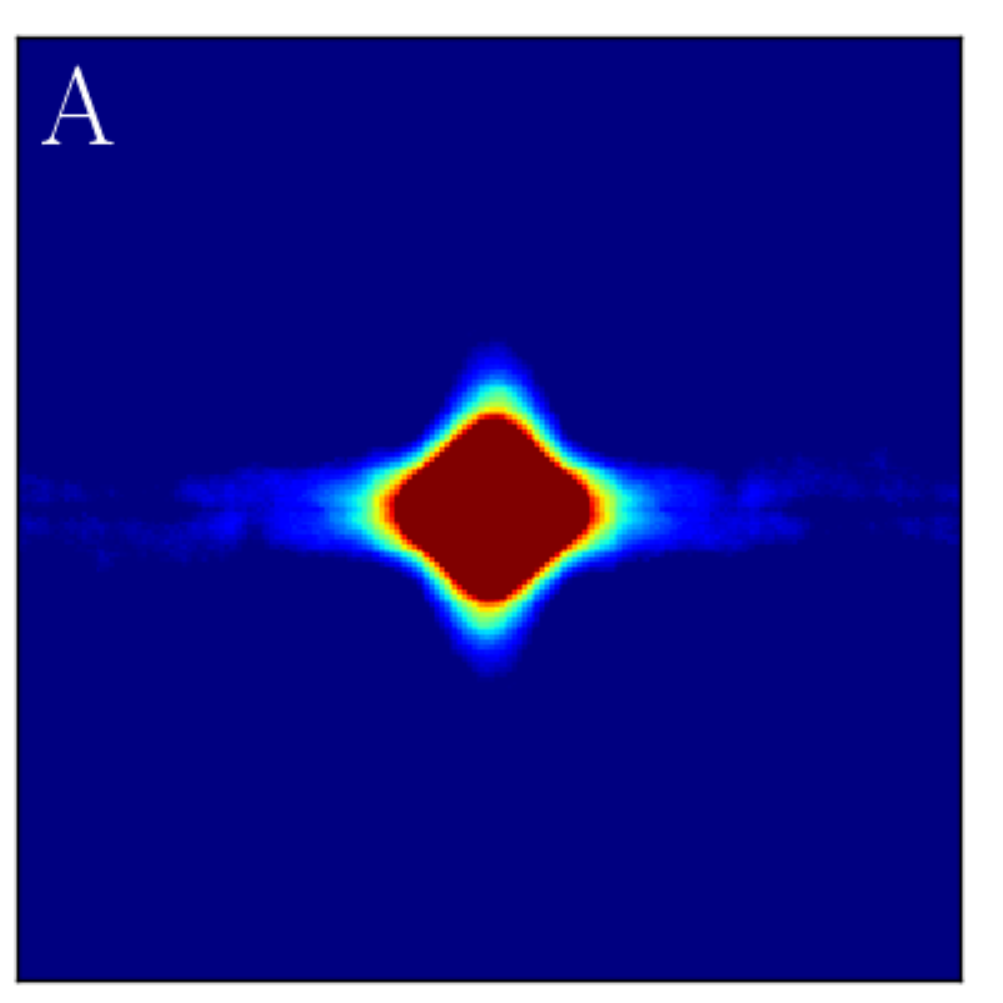}\includegraphics[width=5cm]{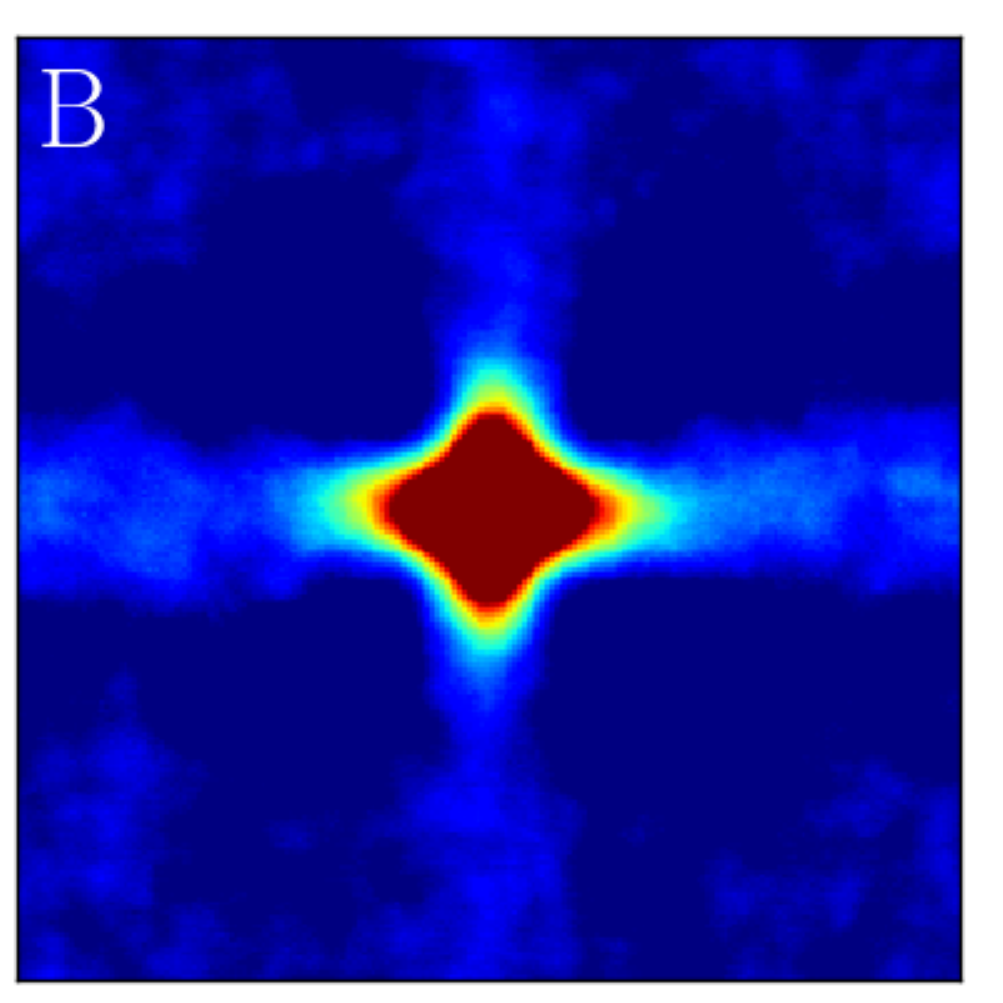}\includegraphics[width=5cm]{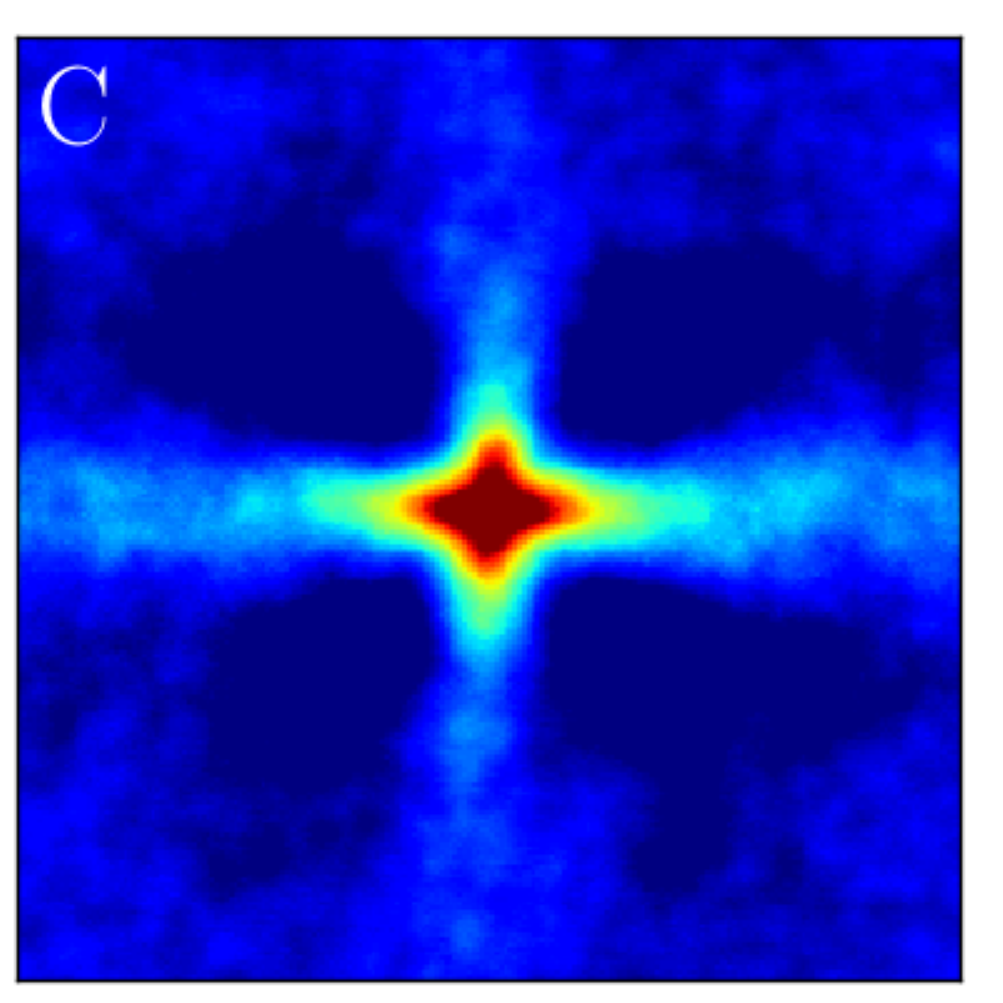}
\par\end{centering}

\begin{centering}
\includegraphics[width=5cm]{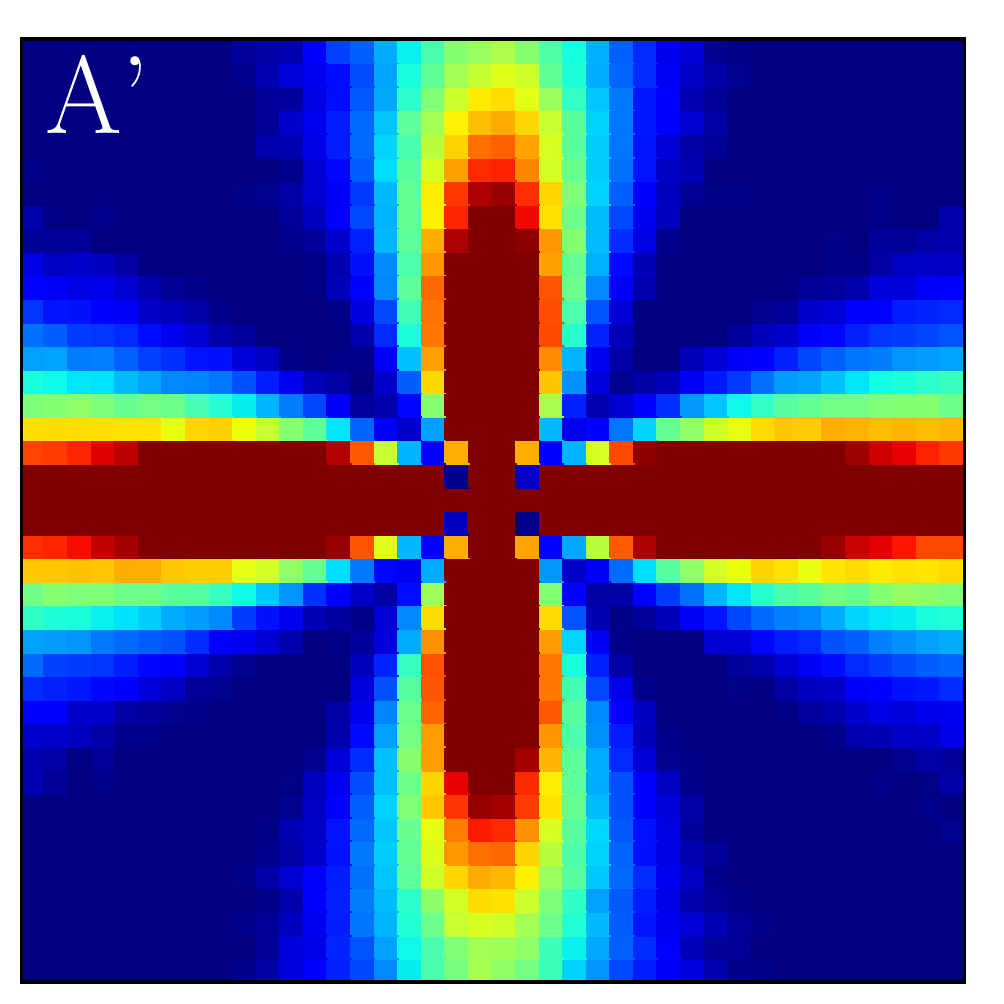}\includegraphics[width=5cm]{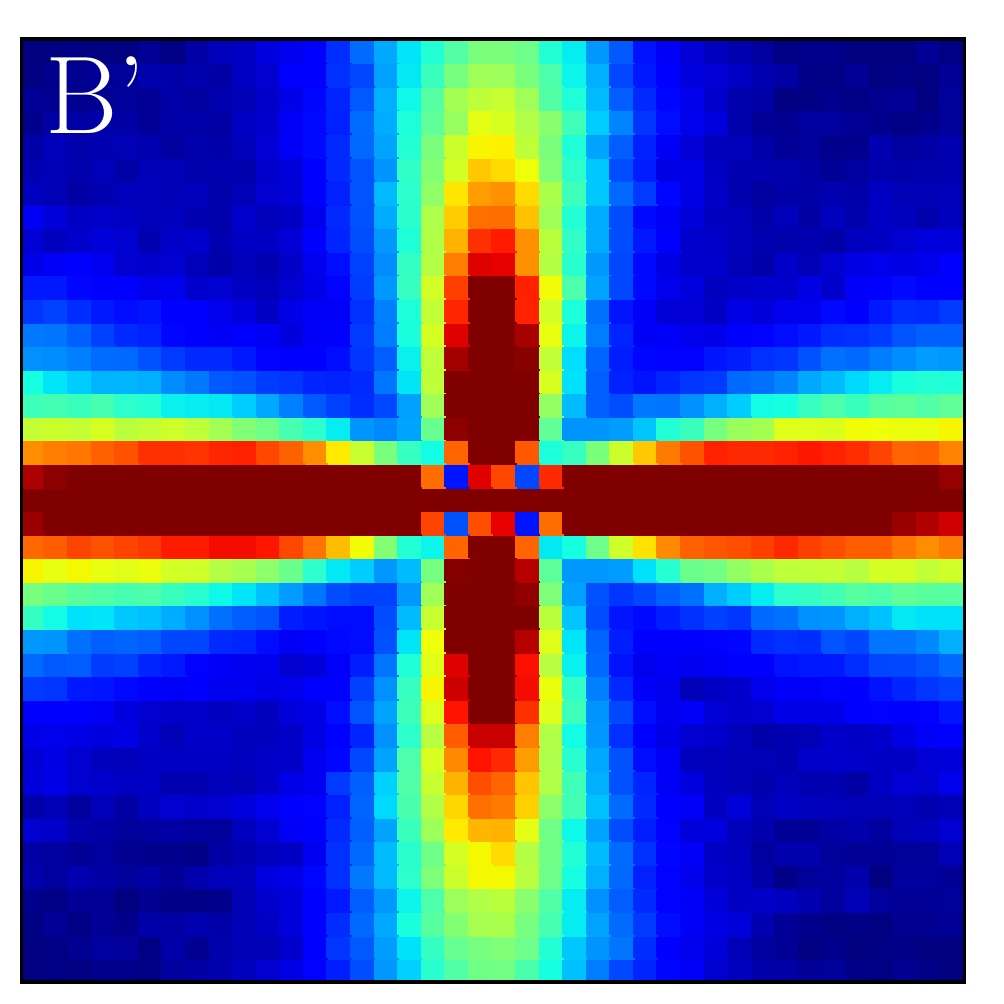}\includegraphics[width=5cm]{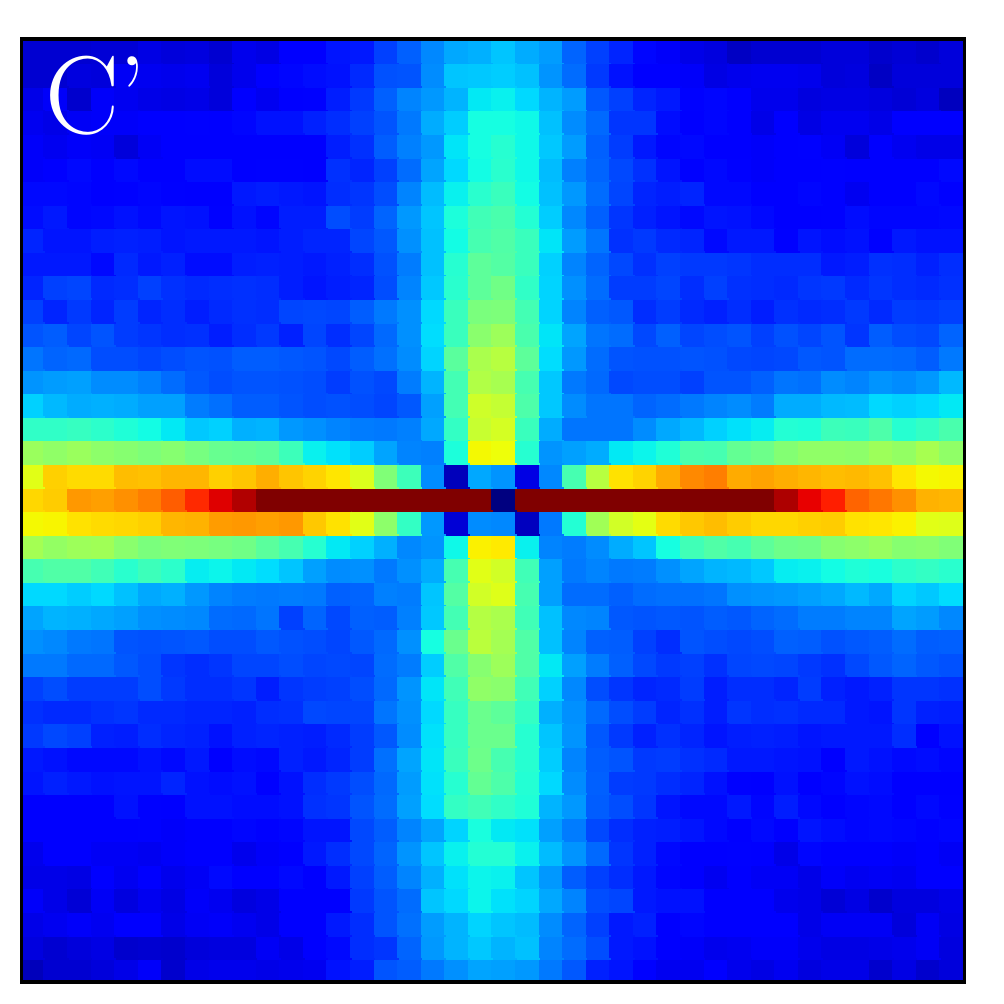}
\par\end{centering}

\caption{\label{fig:PlCorrMEDIUM}Colour maps of the plastic correlator $\mathcal{C}_{2}$
at the intermediate shear rate $\dot{\gamma}=10^{-4}$. Refer to Fig.\ref{fig:PlCorrLOW}
for the rest of the caption.}
\end{figure*}

\begin{figure*}
\begin{centering}
\includegraphics[width=5cm]{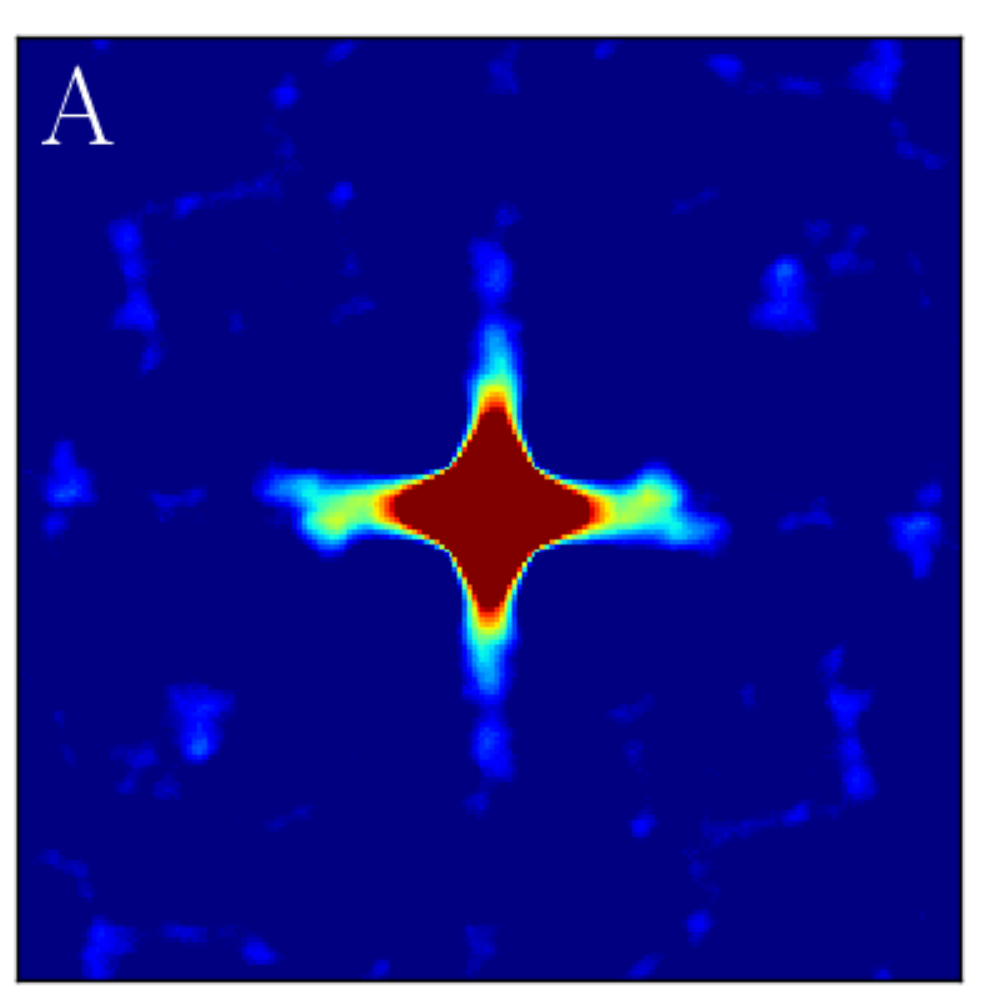}\includegraphics[width=5cm]{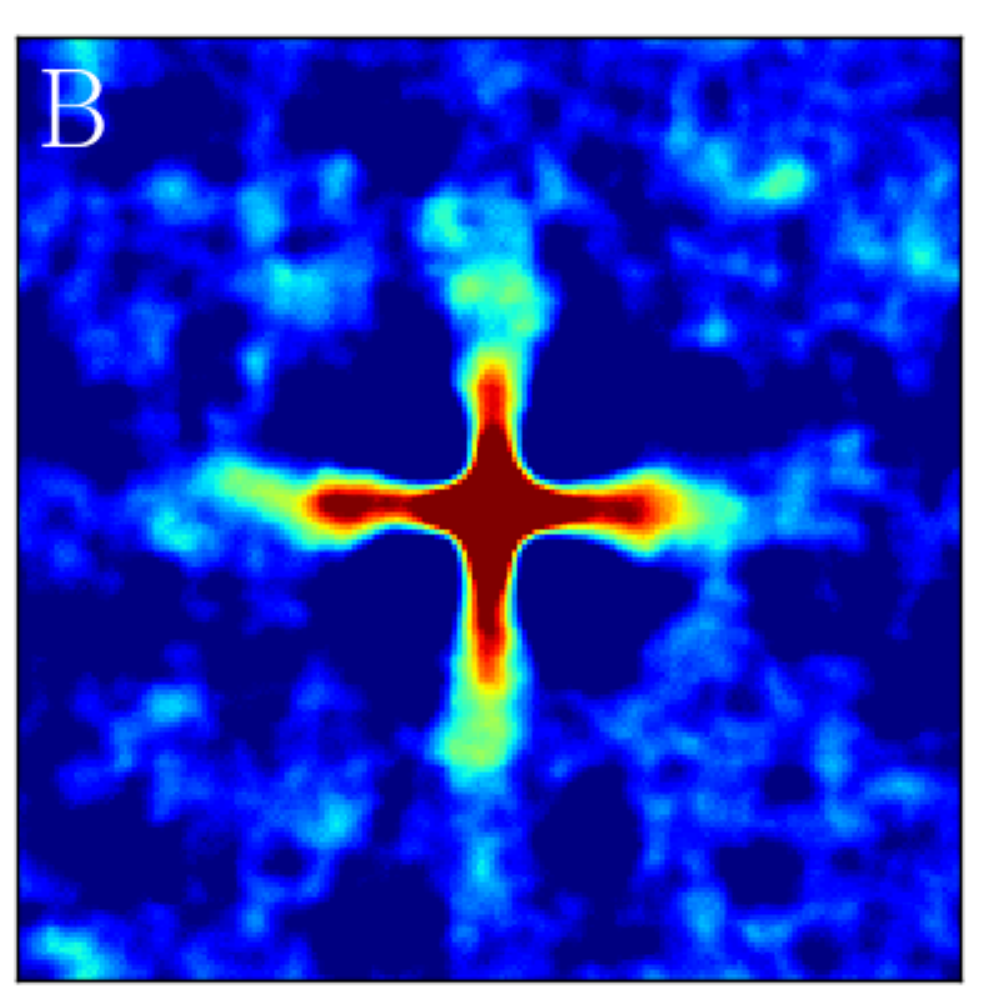}\includegraphics[width=5cm]{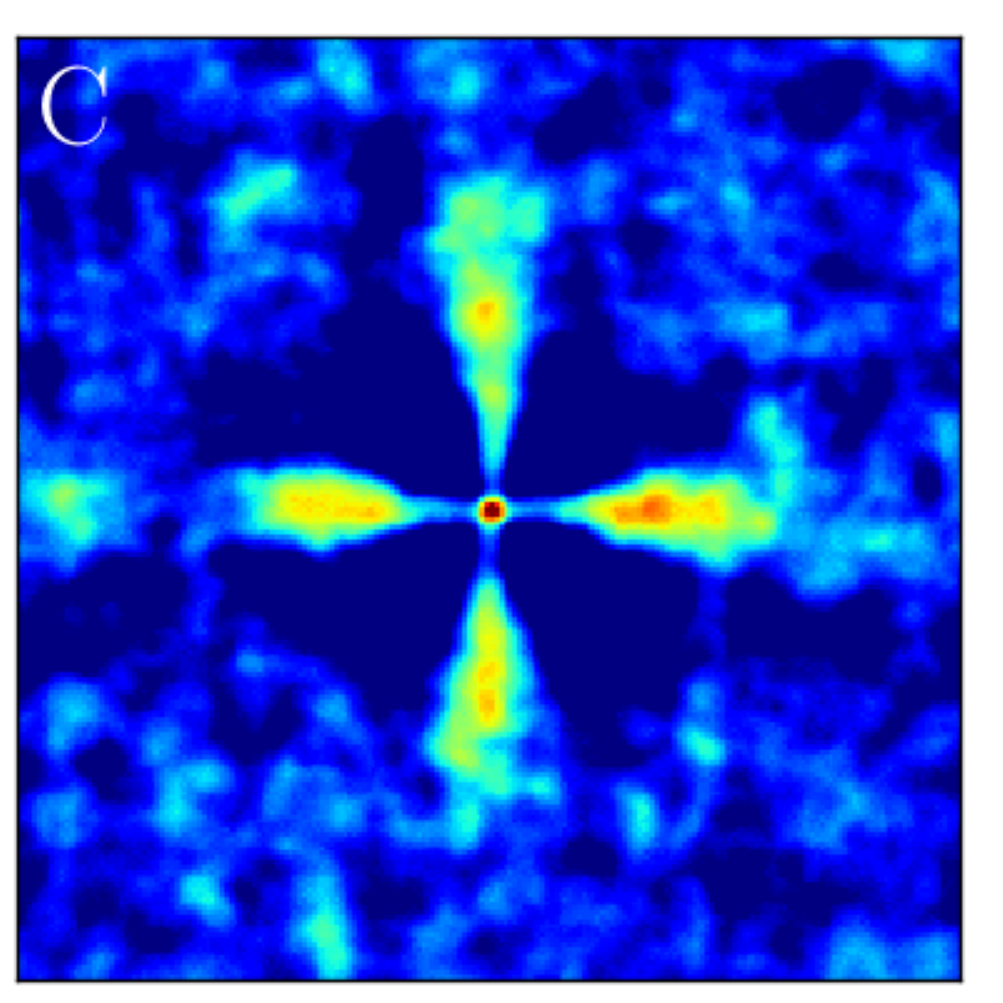}
\par\end{centering}

\begin{centering}
\includegraphics[width=5cm]{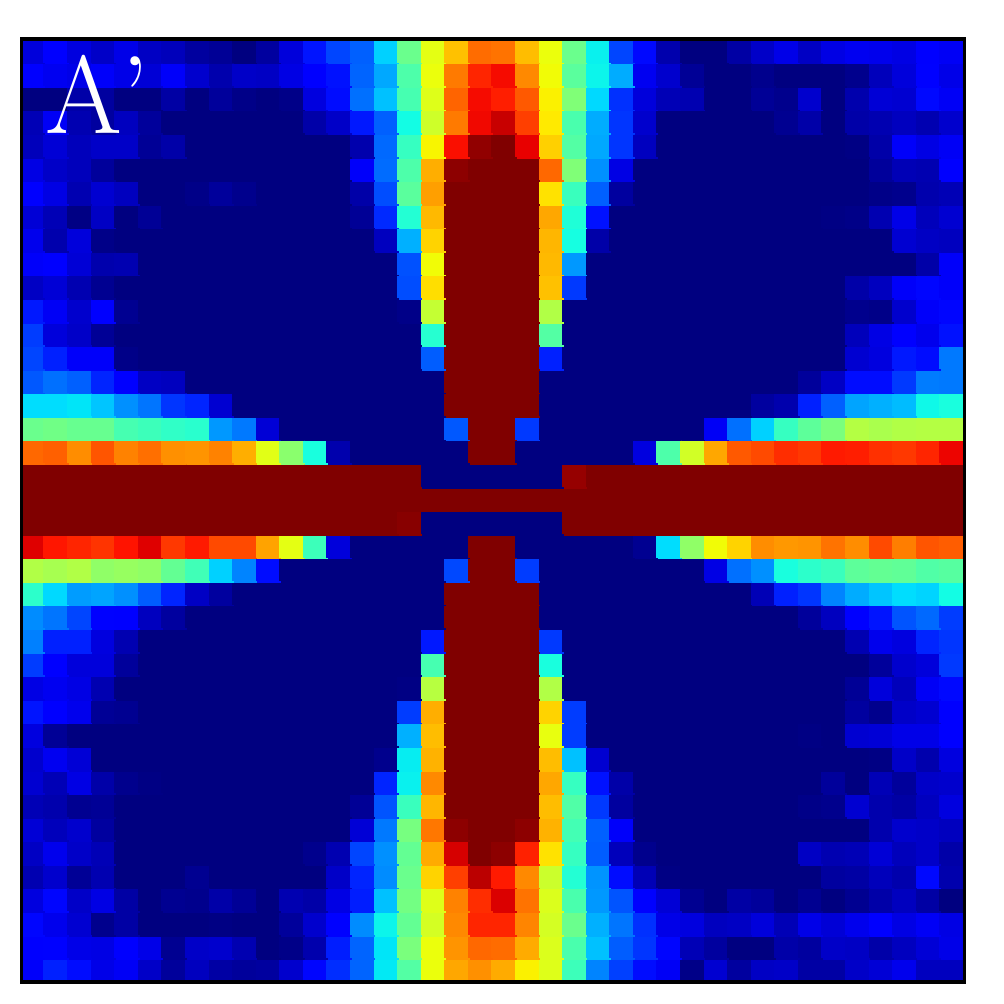}\includegraphics[width=5cm]{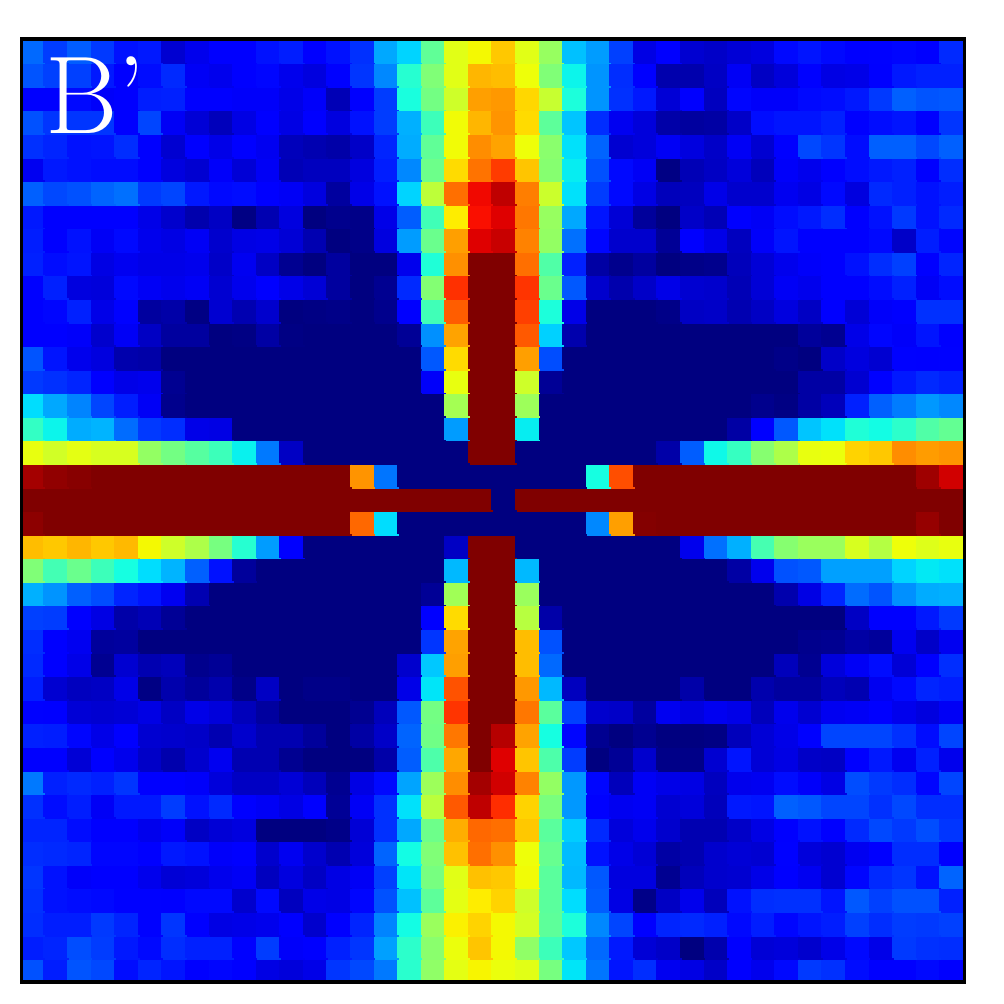}\includegraphics[width=5cm]{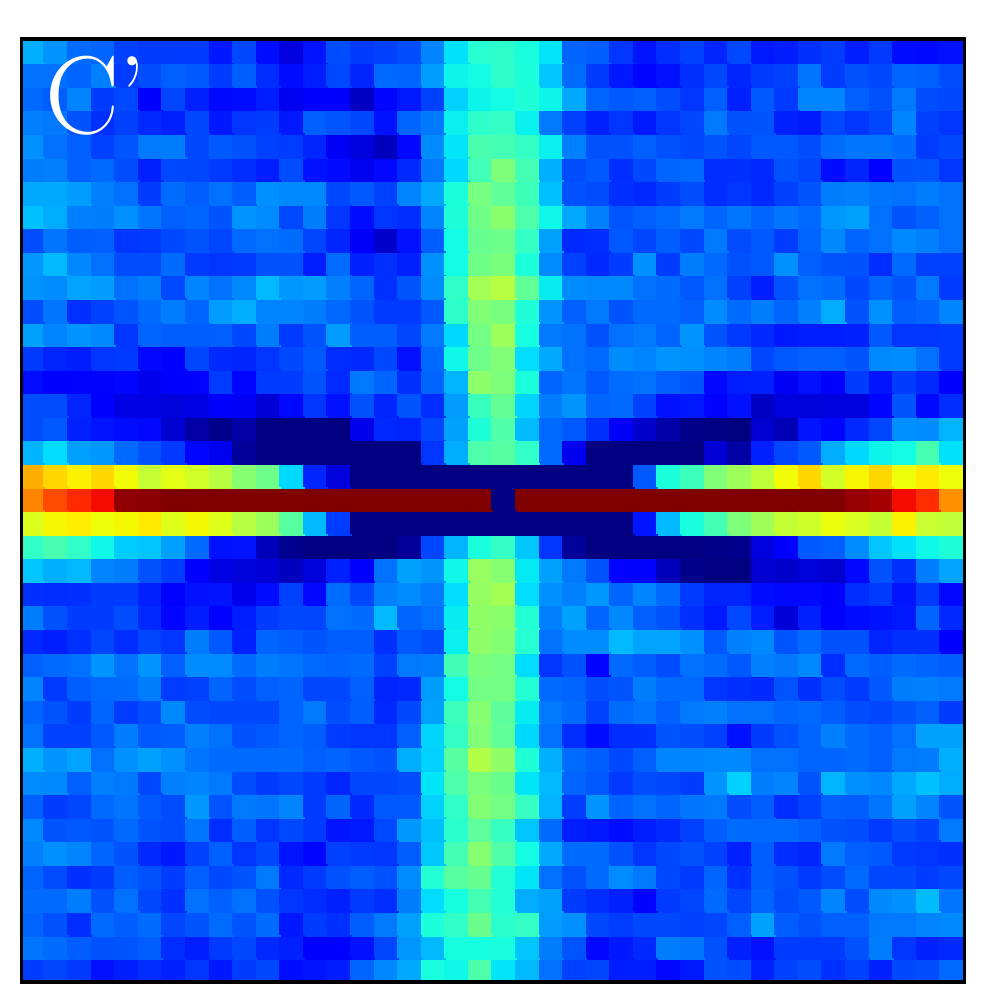}
\par\end{centering}

\centering{}\caption{\label{fig:PlCorrHIGH}Colour maps of the plastic correlator $\mathcal{C}_{2}$
at the moderately high shear rate $\dot{\gamma}=10^{-3}$. Refer to
Fig.\ref{fig:PlCorrLOW} for the rest of the caption. Note that, for
the bottom row, we have used a sightly smaller lower bound for the
colour code, $-1.5\cdot10^{-3}$ instead of $-5\cdot10^{-4}$ in the
other cases.}
\end{figure*}

The plastic correlations obtained in the atomistic simulations are
shown in Fig.\ref{fig:PlCorrLOW}, \ref{fig:PlCorrMEDIUM}, and \ref{fig:PlCorrHIGH}
at different time lags for three distinct shear rates: $\dot{\gamma}=10^{-5}$, $\dot{\gamma}=10^{-4}$, and $\dot{\gamma}=10^{-3}$. The counterparts for the coarse-grained simulations
are presented directly opposite to them so as to allow an easy comparison,
but they will only be discussed below in section \ref{sub:Successes-and-limitations}.

The presence of a spatial structure in the correlations is manifest,
which is strong evidence that plastic rearrangements are indeed interdependent,
and not fully isolated events. The positive correlations in the streamwise
and crosswise directions are strongly reminiscent of the positive
lobes of the elastic propagator $\mathcal{G}$, which supports the
idea of interactions via an elastic coupling. In diagonal directions,
there tend to be anticorrelations.
The (anti)correlations decay gradually, over approximately the same (absolute)
time scale as the autocorrelation function, i.e., their value at the
origin. These features are common to the various shear rates studied here.

\begin{figure}[!ht]
\begin{centering}
%\subfloat[MD simulations]{\begin{centering}
\includegraphics[width=7cm]{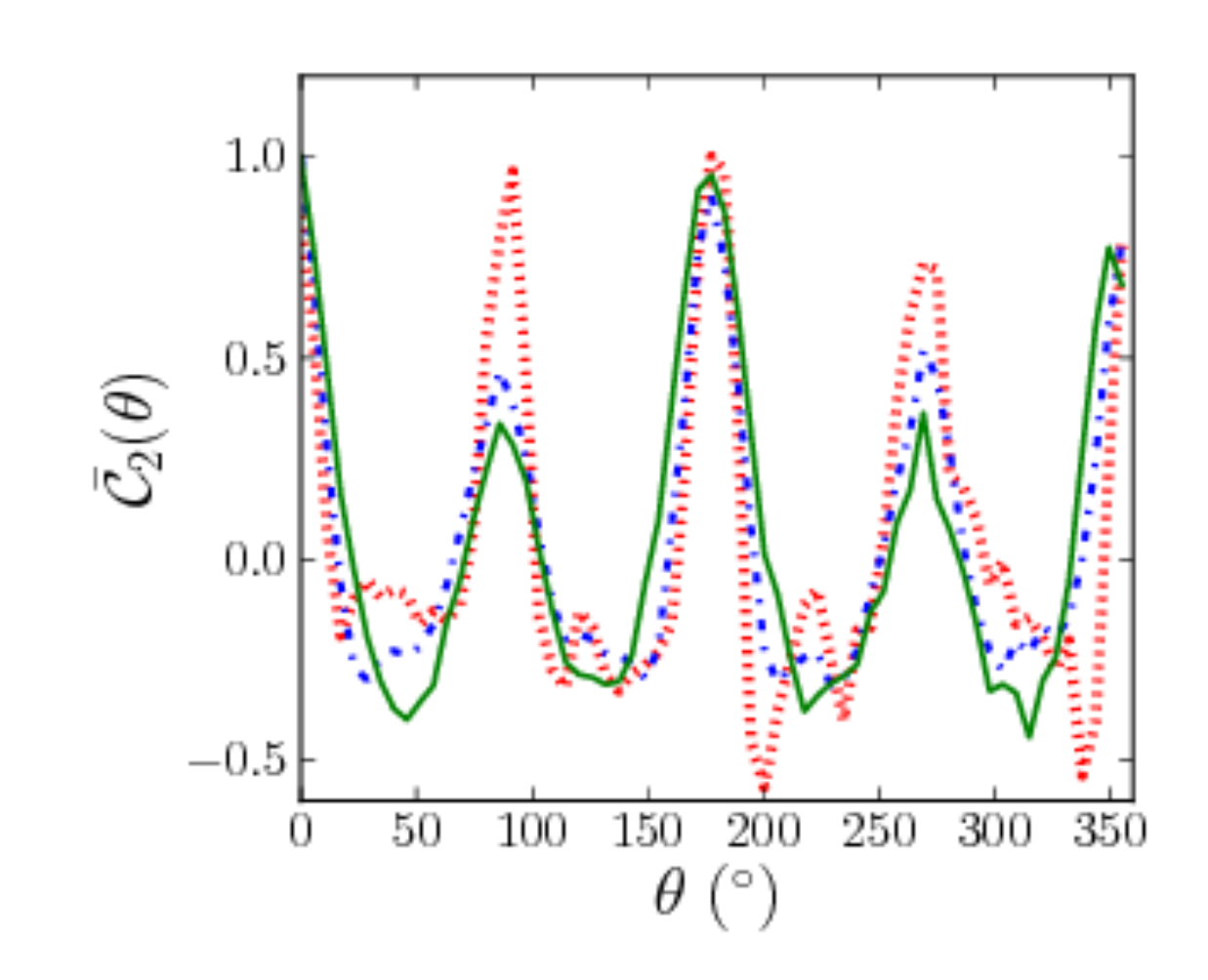}
%\par\end{centering}
\par\end{centering}

\begin{centering}
%\subfloat[Coarse-grained model]{\begin{centering}
\includegraphics[width=7cm]{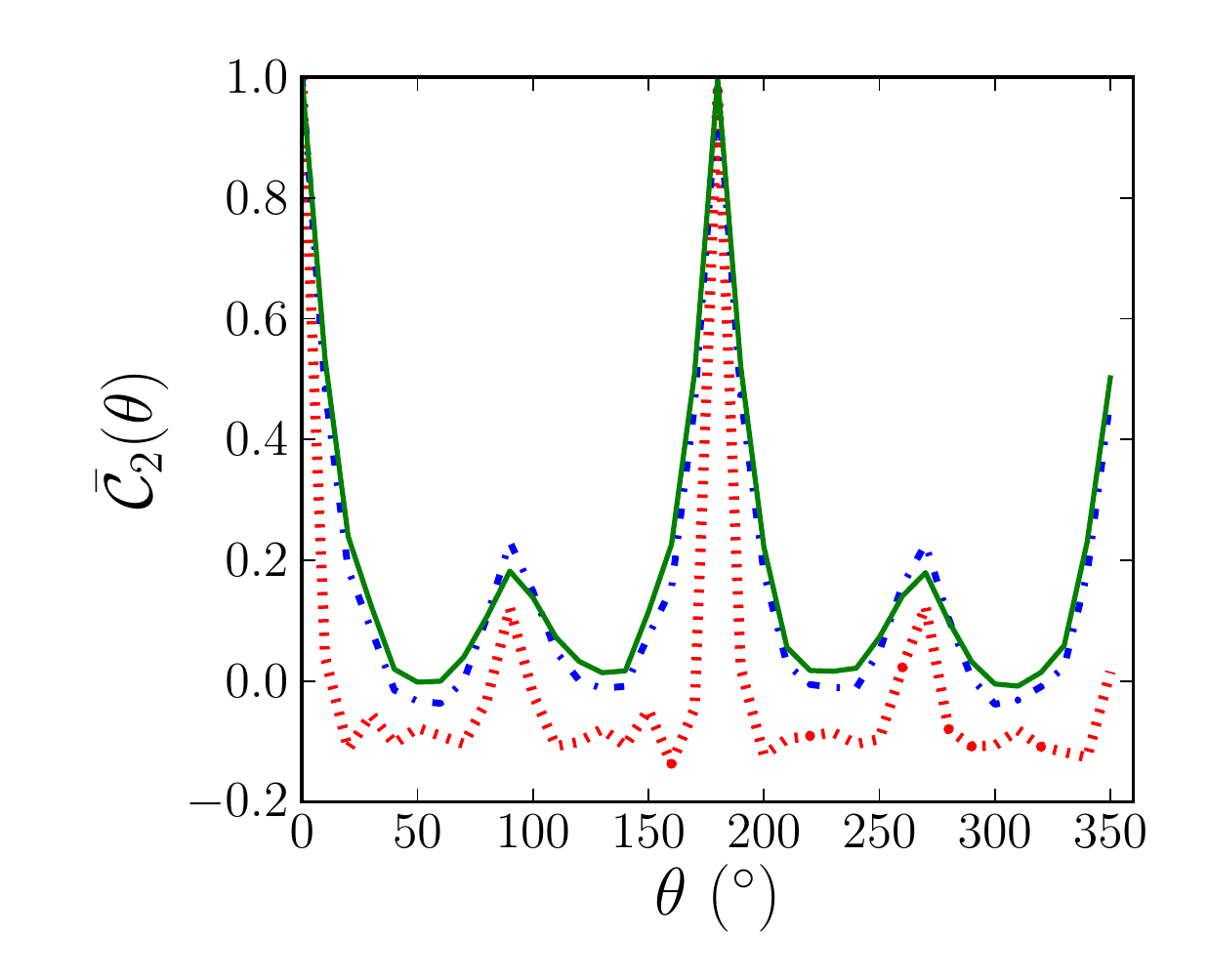}
%\par\end{centering}
\par\end{centering}

\caption{\label{fig:angular_PlCorr}Angular dependence of the correlations
$\bar{\mathcal{C}}_{2}\left(\theta\right)\equiv\alpha\int_{0}^{\nicefrac{L}{2}}\mathcal{C}_{2}\left(r,\theta;\Delta t\right)dr$,
where $L$ is the system size and $\Delta t=20$ is the time lag,
at shear rates (\emph{solid green})$\dot{\gamma}=10^{-5}$, (\emph{dashdotted
blue line})$\dot{\gamma}=10^{-4}$, and (\emph{dotted red line})$\dot{\gamma}=10^{-3}$.
The prefactor $\alpha$ is chosen such that $\bar{\mathcal{C}}_{2}\left(\theta\right)$
has a maximum of 1. Top panel: results from MD simulations. Bottom panel: results from the coarse-grained model.}
\end{figure}

A closer investigation of the plots shows that the decay time tends
to decrease with increasing shear rate, thereby reflecting the shear-induced
decorrelation of the system, with sequences of correlated events being cutoff by the deformation. Moreover, while the streamwise and crosswise
lobes are hardly distinguishable at high shear rates, at lower shear
rates there is clearly an asymmetry between them. The propensity to
shear localisation of the plastic activity is therefore enhanced at
lower shear rates. This is more visible in Fig.\ref{fig:angular_PlCorr}(top panel),
where the correlations are integrated along the radial direction in
different directions. An enhanced propensity
to shear localisation, or, more generally, flow heterogeneities, with decreasing shear rates has already been reported in the literature \cite{Picard2005,Martens2012,Divoux2011}, although, here, some artifact associated with the use of
periodic boundary conditions and finite size effects cannot be excluded\cite{Chattoraj2013}.
An additional effect of the shear rate is that the anticorrelated
lobes in the diagonal directions appear stronger at higher shear rates.

To assess the strength of the correlations, that is, to what extent they deviate from a 
random distribution of plastic events, we compare the probabilities that two plastic 
events, separated by a distance $\Delta r$ and a time lag $\Delta t$, are aligned, on the one hand, along
the velocity gradient direction $\boldsymbol{e_\perp}$  and, on the other hand, along the diagonal direction $\boldsymbol{e_{diag}}$ with respect to the macroscopic shear. We observe an enhancement of the probabilities of streamwise alignment (versus diagonal alignement) by about 10\% to 20\%. Details are provided in Fig.\ref{DirProbEnh} in the Appendix.

\begin{figure}[!ht]
\begin{centering}
%\subfloat[MD simulations]{\begin{centering}
\includegraphics[width=6.8cm]{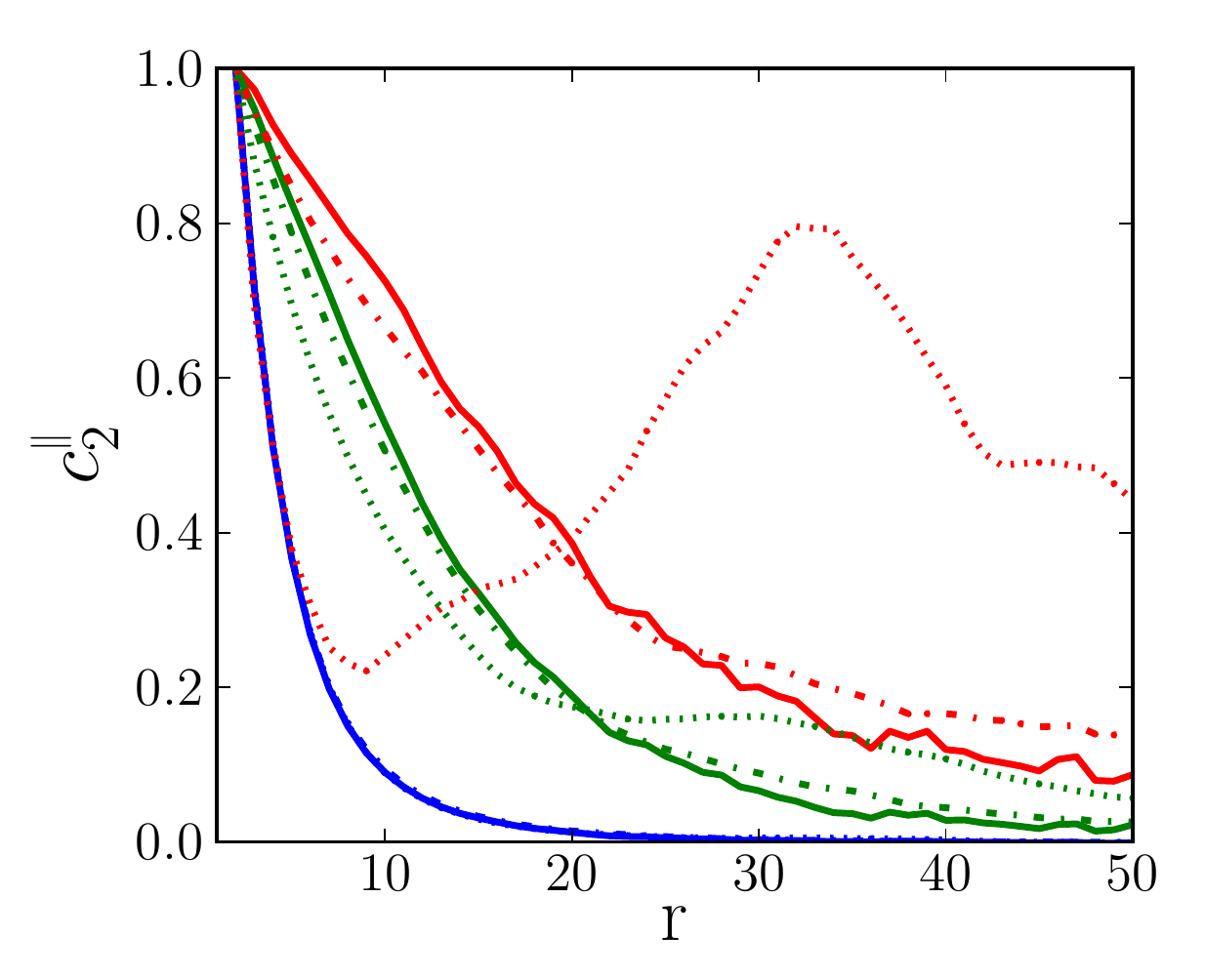}
%\par\end{centering}
\par

%\subfloat[Coarse-grained model]{\begin{centering}
\includegraphics[width=6.8cm]{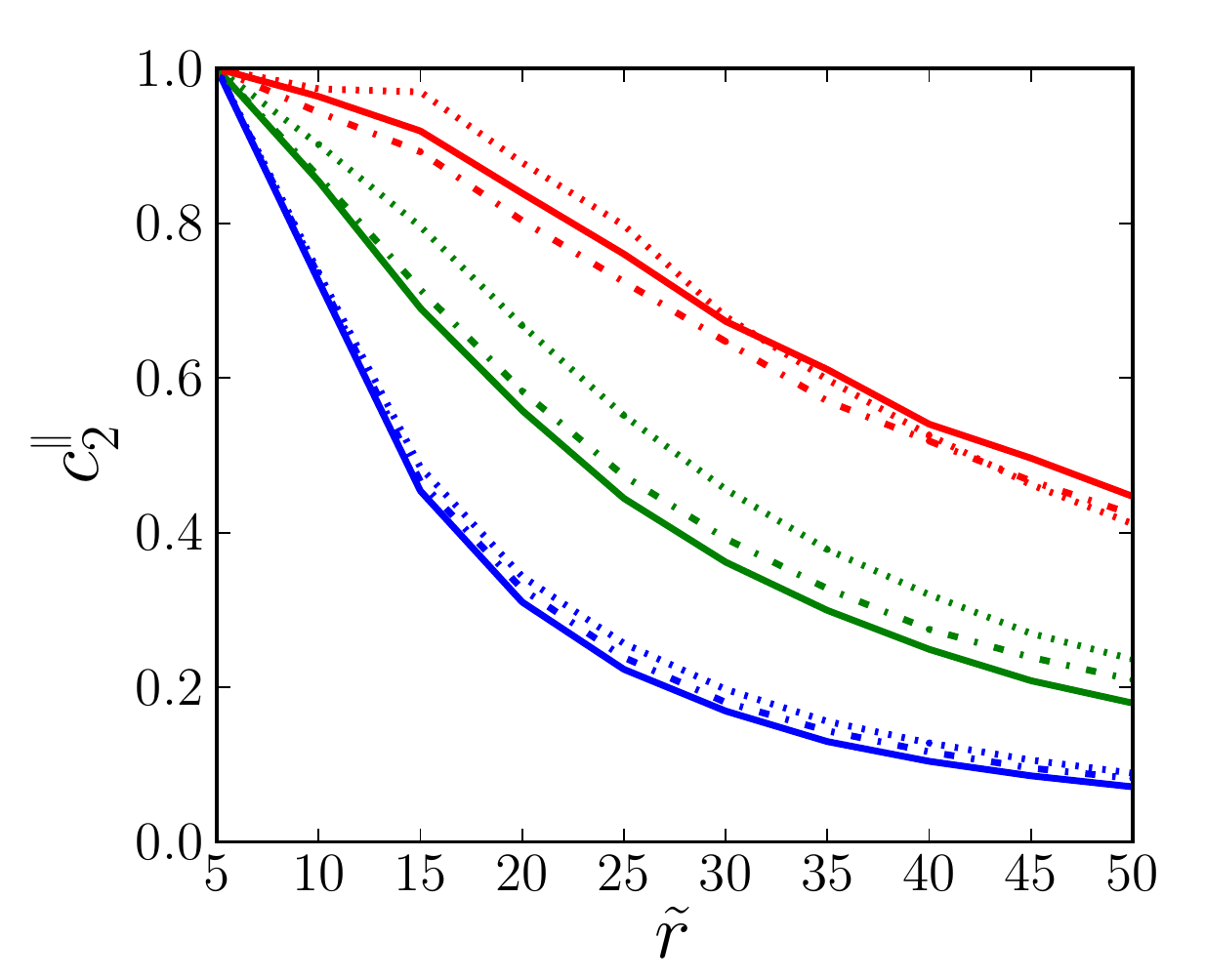}
%\par\end{centering}
\par\end{centering}

\caption{\label{fig:CorrAlongFlow}Correlations along the stream
direction, 
$c_2^{\parallel}(r,\Delta t) \equiv \mathcal{C}_2(r,\Delta t) / \mathcal{C}_2( \epsilon, \Delta t)$,
where the postfactor rescales the correlations to unity close to the origin, $\epsilon=2$ (5 for panel (b)). Data are shown for various time lags: (\emph{blue}) $\Delta t=0$ (1 for bottom panel ),
(\emph{green}) $\Delta t=8$, (\emph{red}) $\Delta t=16$, and for the different shear rates: (\emph{solid line}) $\dot\gamma=10^{-5}$,
 (\emph{dash-dotted line}) $\dot\gamma=10^{-4}$, and  (\emph{dotted line}) $\dot\gamma=10^{-3}$.  Top panel: results from MD simulations. Bottom panel: results from the coarse-grained model.
 To allow direct comparison, we have set the size of one coarse-grained block to $\tilde{r}=5$.
 To reduce the statistical noise, we have averaged the correlations over the three streamlines closest to the origin for the MD simulations.
}

\end{figure}

We now turn to the spatial extent of the correlations. In the top panel of Fig.\ref{fig:CorrAlongFlow}, we show how they
decay along the flow direction, for distinct time lags. The decay, which is not purely exponential, depends only weakly on the shear rate, except at long time lags. Besides, it spreads over larger and larger distances as the time lag
is increased; it should however be noted that the correlations 
have been rescaled so as to be equal to unity close to the origin at all time lags, so that a slower spatial decay does not necessarily imply a larger
\emph{absolute} value far from the origin. This rescaling also entails that small fluctuations will be magnified when the correlations near the origin are small, e.g., for the long time lag $\Delta t=20$, notably in the moderately high shear rate case.

At this stage, we should mention a very recent study by
Varnik and co-workers \cite{Varnik2014}, who reported
that the spatial decay of the $D_{min}^{2}$ correlations
was highly contingent on the specific implementation of the friction
force in the equations of motion
\footnote{Note that, although these researchers have computed nominally
 {}``static'' correlations, that is to say, at $\Delta t=0$,
the time $\delta t$ which they used to compute $D_{min}^{2}$
is very large, so that their data actually correspond to an integral
of our dynamical correlations $\mathcal{C}_{2}\left(\Delta r,\Delta t\right)$
over a wide range of time lags $\Delta t$.)}. 
More precisely, only a friction force based on the relative velocity of a particle with respect to
its neighbours ({}``contact dynamics'') could reproduce the power
law decay observed in experiments on colloidal suspensions and immersed
granular matter, whereas a mean-field dissipation scheme predicted
a faster, exponential decay. The effect of the specific
implementation of the frictional force has been the subject of a wider debate:
Tighe et al.\cite{Tighe2010}, for instance, reported that using a friction term
based on relative particle displacements is key to finding suitable
correlation functions in the vicinity of the jamming point, while
Vagberg et al.\cite{Vagberg2013} claimed that a critical behaviour
is found with both schemes. Here, we have used a mean-field friction force; accordingly, some quantitative discrepancies
may be expected between the extent of the correlations that we have found and those measured in the experimental setups of Ref.
\cite{Chikkadi2012a,Varnik2014}. However, our choice of friction force is, arguably, the more adequate one for confined two-dimensional 
geometries in which particles slide along a fixed plate, for instance, bubble rafts confined in between parallel glass plates \cite{Debregeas2001}.

\newpage{}

\newpage{}

\newpage{}

\subsection{Successes and limitations of the coarse-grained model\label{sub:Successes-and-limitations}}

In Section \ref{sub:Model_vs_MD_macro}, we have seen that the coarse-grained
model gives a rather satisfactory description of the macroscopic properties,
as well as the local ones. Here, we enquire how well it fares with
respect to the full set of spatiotemporal correlations. As shown in Figs.\ref{fig:PlCorrLOW} through \ref{fig:PlCorrHIGH},
the correlation maps for the two models do bear some resemblance,
but closer inspection reveals quantitative differences.

Among the satisfactory aspects, the coarse-grained model also indicates
a decay of the correlations with absolute time and correlations display
a four-fold angular symmetry. By comparing the top and bottom panels of Fig.\ref{fig:angular_PlCorr}, we find reasonable
agreement for the angular dependence of the correlations. One should however admit that excessive
correlations are predicted along the flow direction, especially in
the near field. This is an artifact associated with the use of a regular
lattice: as the frame is deformed, the positive lobe of the elastic
propagator in the flow direction remains aligned with one axis of
the lattice, while the alignment of the perpendicular lobe with the
other axis is lost.
Moreover, the coarse-grained simulations are able to describe the anti-correlated
lobes in the diagonal directions and their enhancement at higher shear
rates.

On the downside, it is obvious that salient features of the plastic
correlations are amiss. This discrepancy is interesting, because it
is a hint that plastic correlations reveal some physical processes
that may otherwise be left unnoticed, and that these processes have
been omitted in the model. 
First, the gradual growth with time of the correlations observed in
MD is in stark contrast with the maximal extent of the correlations
at vanishing time lag in the coarse-grained model, as can be seen in Fig.\ref{fig:CorrAlongFlow}(bottom panel). 
Indeed, within a short time lag $\Delta t<1$, the mesoscopic model builds up correlations
over a characteristic distance of order 20, while these quasi-instantaneous correlations
do not extend beyond a few unit lengths in the MD simulation.
This indicates that the MD correlations do not grow because more and more shear stress
is redistributed as the rearrangement proceeds, but because shear
waves need a finite time to propagate (whereas instantaneous equilibration
was assumed in the model). In other words, the acoustic delay for the propagation
of strain-waves within an avalanche slows down the emergence of spatial correlations. Indeed, the initial growth of the correlations 
is consistent with the propagation of shear waves at the transverse sound velocity (of the undamped system), viz.,
$c_{t}=\sqrt{\nicefrac{\mu}{\rho}}\simeq4$. The gradual expansion of the strain field created by a plastic event is studied in more details
in the companion paper, Ref.\cite{Puosi2014}, and in Ref.\cite{Idema2013,Chattoraj2013}. Note that, in this last reference, the authors also observed some advanced frontline moving at the longitudinal sound velocity $c_l > c_t$.

The second major difference lies in the spatial extent of the correlations,
which is much larger in the coarse-grained approach. Had a frictional force based on \emph{relative} velocities been used, MD may have yielded larger correlations, as suggested by ref.\cite{Varnik2014}.
However,
the large deviation between the predictions of
the atomistic and coarse-grained models does point to an additional
source of discrepancy.
We believe that the underestimation of structural disorder in the
coarse-grained model is at the core of the divergence. Indeed, broadening
the distribution of energy barriers in the model results in somewhat
shorter correlations, at the expense of a poorer fitting of the macroscopic flow properties by
our essentially one-parameter model.  Of probably equal relevance
is the use of an 'ideal' elastic propagator. This propagator
describes stress redistribution in a perfectly uniform elastic
medium. Such a description is justified \emph{on average},
but is inaccurate for a specific plastic event\cite{Puosi2014}, because
elastic heterogeneities in the surrounding medium, i.e., the spatial
variations of the local elastic constants, induce deviations from
the ideal case. The insufficient account of structural disorder in the model is also reflected in the vastly overestimated anisotropy of
the correlations that it predicts, as measured by the directional probability enhancement (see the bottom panel of Fig.\ref{DirProbEnh} in the Appendix).

\section{Summary and outlook}

In conclusion, we have reported  numerical simulations that confirm
the basic flow scenario for amorphous solids, based on swift localised
rearrangements embedded in an elastic matrix and interacting via an
elastic deformation field. A coarse-grained model based on this simple scenario 
satisfactorily reproduces the measured flow curve, the surface density of simultaneous plastic events,
and the decay of the stress autocorrelation function.

To obtain full insight into the dynamical organisation of flow heterogeneities, we have probed the spatio-temporal 
correlations between plastic events, and their dependence on the shear rate.
As already reported in the literature, these correlations are perceivably anisotropic and exhibit
the four-fold angular symmetry characteristic of the elastic propagator. These correlations spread approximately at the 
transverse sound velocity before fading away. Varying the shear rate only brings on small changes to the general picture: at 
higher shear rates, the near-field anticorrelations along the diagonal directions seem to be slightly more pronounced, and the velocity 
and velocity gradient directions are more symmetric. Besides, the spatial extent of the correlations tends to decrease with increasing shear rate.

A coarse-grained model is able to describe the observed symmetries of the correlations, 
but fails to reproduce their emergence in time, owing to the neglect of the finite shear wave velocity in the model.
In addition, the model vastly overestimates the anisotropy in the correlations, thereby pointing to the 
underestimation of structural disorder in the system, at least partly because of the use of an ideal elastic propagator.
These two flaws are not specific to the model used here, but a general deficiency of all approaches of this type\cite{Rodney2011}. Consequently, should one aim for a proper description of these correlations, these missing physical aspects will need to be incorporated into the models. To what extent they will alter 
the macroscopic flow properties predicted by the models, for instance the variable propensity to shear localisation, remains an open question. More generally, it seems likely that in the future the study of plastic correlations, as described in Ref. \cite{Chattoraj2013,Benzi2014} or in the present work, will become a powerful tool for comparison between models of various types and between models and experiments in systems in which the corresponding observables are experimentally accessible.

\textbf{Acknowledgments}
  The simulations were carried out using the  LAMMPS molecular dynamics software\footnote{http://lammps.sandia.gov}. JLB is supported by Institut Universitaire de France and by grant ERC-2011-ADG20110209.  JR acknowledges support from Universit\'e Joseph Fourier and from Institut Laue Langevin during a stay in Grenoble.
%
%\end{acknowledgments}

\appendix

\section*{Quantification of the anisotropy of the correlations}

\begin{figure}[ht]
\begin{centering}
%\subfloat[MD simulations]
{\includegraphics[width=7cm]{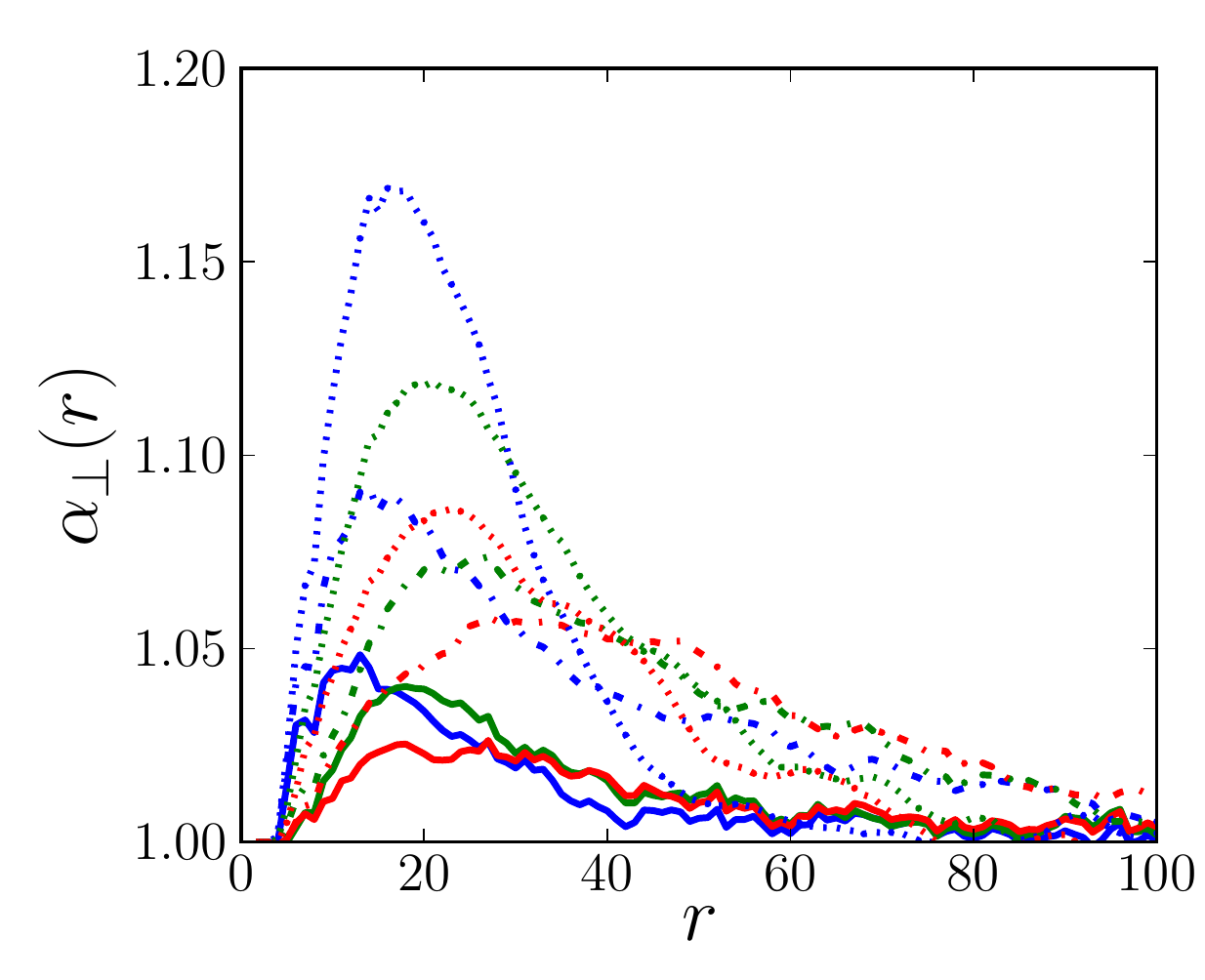}

}
\par\end{centering}

\begin{centering}
%\subfloat[Coarse-grained model]
{\includegraphics[width=7cm]{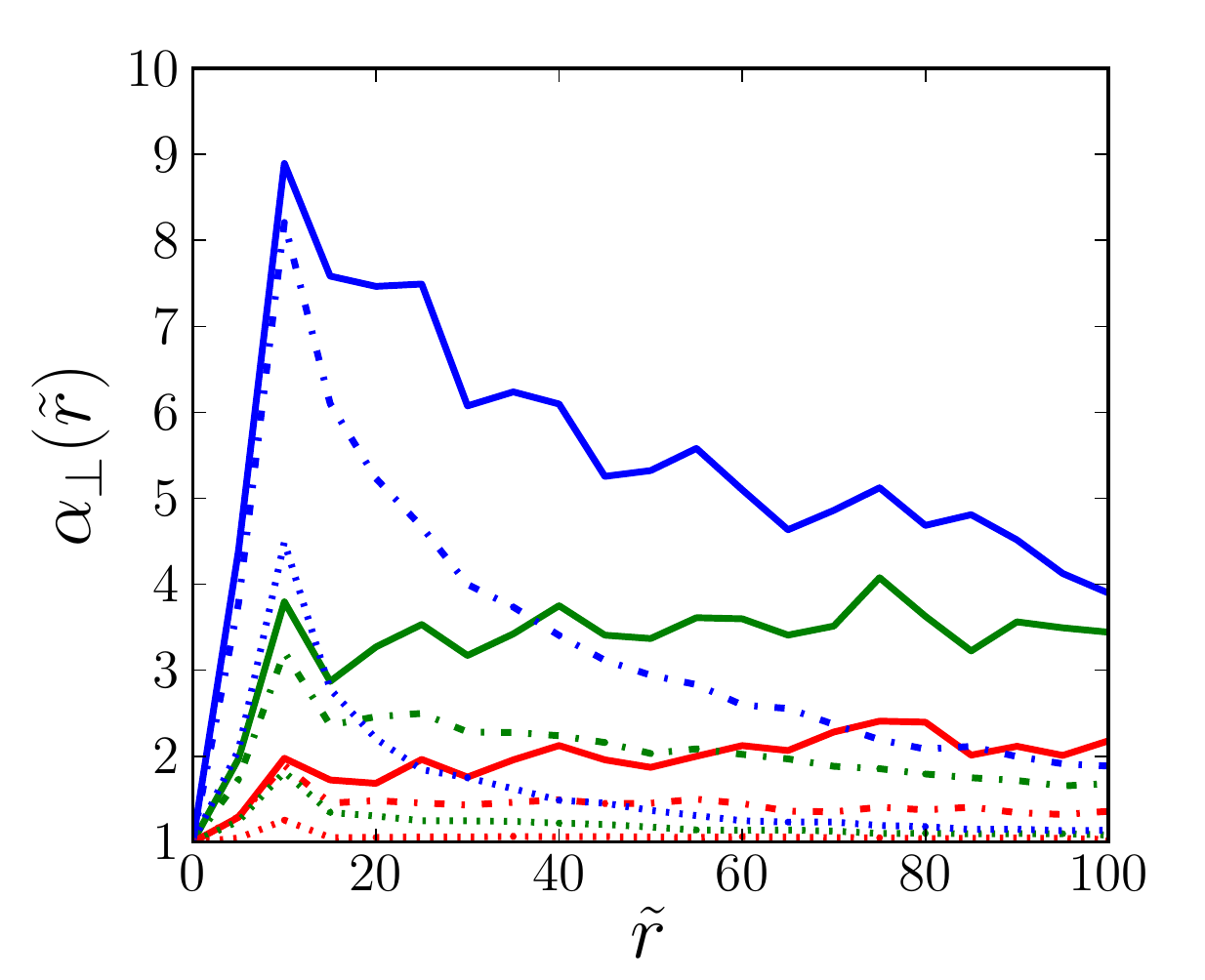}}
\par\end{centering}

\caption{\label{DirProbEnh}Directional probability enhancement factor$\alpha_\perp(\Delta r,\Delta t,\dot{\gamma})$
for the different shear rates: (\emph{solid line}) $\dot{\gamma}=10^{-5}$,
(\emph{dash-dotted line}) $\dot{\gamma}=10^{-4}$, and (\emph{dotted
line}) $\dot{\gamma}=10^{-3}$, and for various lag times. \emph{Top
panel}: (\emph{blue}) $\Delta t=4$ (1 for panel (b)), (\emph{green})
$\Delta t=12$, (\emph{red}) $\Delta t=20$. \emph{Bottom panel}:
(\emph{blue}) $\Delta t=0$ (1 for panel (b)), (\emph{green}) $\Delta t=8$,
(\emph{red}) $\Delta t=16$. To allow direct comparison, we have set
the size of one coarse-grained block to $\tilde{r}=5$.}

\end{figure}

To assess the strength of the anisotropy in the plastic correlations, we compute
the directional probability enhancement factor $\alpha_\perp$,
\emph{viz.},
\begin{eqnarray}
\alpha_{\perp}(\Delta r,\Delta t,\dot{\gamma}) & \equiv & \frac{\langle D_{min}^{2}({\bf r},t)\cdot D_{min}^{2}({\bf r}+\Delta r\boldsymbol{e_{\perp}},t+\Delta t)\rangle}{\langle D_{min}^{2}({\bf r},t)\cdot D_{min}^{2}({\bf r}+\Delta r\boldsymbol{e_{diag}},t+\Delta t)\rangle},
\end{eqnarray}
where $\boldsymbol{e_{\perp}}$ and $\boldsymbol{e_{diag}}$
are the velocity gradient and diagonal directions, respectively.
This factor measures the ratio of the probabilities that two plastic
events, separated by $\Delta r$ in distance and $\Delta t$ in time,
will be aligned along the velocity gradient direction
versus diagonally. 
Figure~\ref{DirProbEnh} compares this enhancement ratio for the MD model and 
the coarse-grained model.

\end{document}